\documentclass[acmsmall, manuscript]{acmart}

\usepackage{soul}
\usepackage{enumitem}
\usepackage{algorithmicx} % not
\usepackage{algorithm}
\usepackage[noend]{algpseudocode}   % not
\usepackage{graphicx}
\usepackage{booktabs}
\usepackage{multirow}
\usepackage{xcolor}
\usepackage[table]{xcolor}
\usepackage{colortbl}
\usepackage{tcolorbox}
\tcbuselibrary{skins}
\usepackage{stfloats}
\usepackage{bm}  
\usepackage{listings}
\usepackage{makecell} 
\usepackage[utf8]{inputenc}
\usepackage[table]{xcolor}
\usepackage{etoolbox}
\usepackage{array}

\usepackage{xcolor}
\usepackage{setspace}
\usepackage{enumitem}
\usepackage{float}

\sethlcolor{yellow!50}

\AtBeginDocument{%
  }

\setcopyright{acmlicensed}
\copyrightyear{2026}
\acmYear{2026}
\acmDOI{XXXXXXX.XXXXXXX}

\acmJournal{TOSEM}
\acmVolume{00}
\acmNumber{00}
\acmArticle{00}
\acmMonth{00}

\begin{document}

% \newpage
% \input{revision/00cover}
% \newpage
% \input{revision02/01AE}
% \clearpage
% \input{revision02/02reviewer1}
% \input{revision02/03reviewer2}
% \input{revision02/04reviewer3}
% \newpage

\title[Enhancing LLM Performance Through Debate]{Enhancing LLM Performance Through Debate: An Empirical Study on Multi-Agent Debate for Coding Tasks}

\author{Yong Jin Chun}
\email{chunyj@uci.edu}
\affiliation{%
\institution{University of California, Irvine}
\city{Irvine}
\state{California}
\country{USA}
}

\author{Qihong Chen}
\email{chenqh@uci.edu}
\affiliation{%
\institution{University of California, Irvine}
\city{Irvine}
\state{California}
\country{USA}
}

\author{Jiawei Li}
\email{jiawl28@uci.edu}
\affiliation{%
\institution{University of California, Irvine}
\city{Irvine}
\state{California}
\country{USA}
}

\author{Iftekhar Ahmed}
\email{iftekha@uci.edu}
\affiliation{%
\institution{University of California, Irvine}
\city{Irvine}
\state{California}
\country{USA}
}

\renewcommand{\shortauthors}{Chun et al.}

\begin{abstract}
Large Language Models (LLMs) have advanced autonomous agents' planning and decision-making, yet they struggle with complex tasks requiring diverse expertise and multi-step reasoning. Multi-Agent Debate (MAD) systems, introduced in NLP research, address this gap by enabling structured debates among LLM-based agents to refine solutions iteratively. MAD promotes divergent thinking through role-specific agents, dynamic interactions, and structured decision-making. Recognizing parallels between Software Engineering (SE) and collaborative human problem-solving, this study investigates MAD’s effectiveness on four coding tasks in SE. We adapt a MAD framework from NLP, analyze agent interactions to assess consensus-building and iterative refinement, and propose two MAD variants that enhance agent debate for coding tasks by addressing the observed weaknesses. Our findings show that structured debate and collaboration improve problem-solving and yield strong performance in some cases, highlighting the collaborative debate synergy between the LLM-agents for coding tasks in SE while identifying areas for exploration.
\end{abstract}





\maketitle

\newcommand{\llmAgents}{{LLM agents}}

\setcounter{table}{0}
\section{Introduction}
\label{section:intro}
% definition of intelligent systems
Intelligent systems that plan their actions to complete a task autonomously based on feedback from the environment~\cite{franklin1996agent, maes1993modeling} have been extensively studied over time. Such systems often lacked the adaptability and decision-making sophistication needed to solve complex tasks~\cite{maes1993modeling}. The recent advent of Large Language Models (LLMs) has significantly improved planning and decision-making capabilities~\cite{hou2024large, wei2022chain, zheng2025towards, du2023improving} by serving as a core component of these intelligent systems. With LLM integration, these systems have advanced to demonstrate improved adaptability, problem-solving skills, and autonomy, leading to their wide adoption in solving real-world problems across diverse domains~\cite{huang2022language, yao2023react, li2024agentsneed}.

% LLM-based systems in SE, and its limitations
However, the performance of these LLM-based systems is still limited when applied to Software Engineering (SE) coding tasks that require a deep understanding of algorithms and programming languages or involve complex, multi-step reasoning processes~\cite{he2024llm, hou2024large, zheng2025towards}. coding tasks also exhibit inherent complexity, as they often involve more than one correct solution~\cite{brooks1987no}. For instance, in code translation, there are multiple valid ways to translate code from one language to another, while subtle syntactic or semantic errors can result in program failures. These characteristics make LLM-based systems prone to subtle errors when applied to coding tasks.
% current LLM-based methods in SE
To address this, studies have explored the integration of tool sets~\cite{yao2023react, xia2024agentless}, different prompting techniques~\cite{madaan2023self, chan2024chateval, ahmed2024automatic}, and fine-tuning techniques~\cite{suh2024empirical, murali2024ai, ding2024semcoder} to improve the adaptability of LLM-based systems for solving coding tasks that require domain knowledge and expertise. However, developing and maintaining advanced tools that are compatible with the rapid evolution of LLMs is challenging~\cite{10.1145/3712005}. In addition, prompting techniques often rely on extensive trial-and-error processes and fine-tuning approaches are constrained by the limited availability of large-scale, high-quality, domain-specific datasets. These limitations highlight the need for LLM-based systems that are both applicable and adaptable in the absence of resources and external tools.

Recent studies in the Natural Language Processing (NLP) domain have introduced \textbf{Multi-Agent Debate (MAD)} systems that incorporate \llmAgents{} with diverse expertise and responsibilities~\cite{li2024improving}, demonstrating effectiveness in solving complex NLP tasks without relying on fine-tuning or external tool usage~\cite{du2023improving, smit2024goingmadlookmultiagent, liang-etal-2024-encouraging}. In MAD systems, agents work toward a shared objective by exchanging information and iteratively refining decisions~\cite{khan2024debating, liang-etal-2024-encouraging}, inspired by human collaboration, where individuals exchange ideas, debate discrepancies, and iteratively refine solutions to reach consensus~\cite{li2024improving}. Notably, these systems integrate both collaboration and structured opposition by incorporating mechanisms such as debate, negotiation, and decision-making to foster divergent thinking and enhance the problem-solving capabilities of \llmAgents{}~\cite{chan2024chateval, chen2024magdi}. This competitive framework forces agents to engage in interactions that drive more robust reasoning and solution refinement~\cite{smit2024goingmadlookmultiagent, du2023improving}.
% By decomposing tasks into smaller subtasks and structuring debate form of collaboration, MAD enables agents to engage in interactions that drive more robust reasoning and solution refinement~\cite{smit2024goingmadlookmultiagent, du2023improving}. Prior work has shown that such collective intelligence enhances divergent thinking~\cite{liang-etal-2024-encouraging}, multi-step reasoning~\cite{du2023improving, smit2024goingmadlookmultiagent}, and decision-making~\cite{chan2024chateval, chen2024magdi} capabilities of \llmAgents{}, leading to improved performance on complex tasks.
In this study, we investigate the effectiveness of MAD for coding tasks, which remains largely unexplored in SE research despite its demonstrated effectiveness in NLP tasks such as question answering and arithmetic reasoning~\cite{du2023improving, smit2024goingmadlookmultiagent, liang-etal-2024-encouraging, Xiong_2023, choi2025debate, estornell2025multi}. In particular, we first evaluate the applicability of MAD from the NLP literature and the effectiveness of debate-based collaboration among LLM agents for coding tasks. This motivates the first research question of our empirical study:

\vspace{5pt}
% addressing and answering each RQs
\noindent\textbf{RQ1: How effective is Multi-Agent Debate when applied to coding tasks?}

%NEW IA
% To answer RQ1, we first implement MAD following the State-Of-The-Art (SOTA) approach proposed in NLP by Liang et al.~\cite{liang-etal-2024-encouraging}. Among existing MAD frameworks~\cite{liang-etal-2024-encouraging, chan2024chateval, du2023improving}, we select Liang et al.~\cite{liang-etal-2024-encouraging} as our MAD baseline as it is the most recent state-of-the-art MAD framework, evaluating both natural language generation and reasoning tasks, which closely matches our task context for coding tasks that involve both natural language and code.
% We then evaluate its applicability and effectiveness on coding tasks, comparing its performance with SOTA techniques from prior SE studies. 

To answer RQ1, we implement the state-of-the-art (SOTA) MAD framework proposed by Liang et al.~\cite{liang-etal-2024-encouraging} as our baseline. Among existing MAD frameworks~\cite{liang-etal-2024-encouraging, chan2024chateval, du2023improving}, we select Liang et al.~\cite{liang-etal-2024-encouraging} as it is the most recent SOTA, evaluated on both natural language generation and mathematical reasoning tasks, which closely matches the context of coding tasks that involve both natural language understanding and logical reasoning. To adapt this NLP-based MAD framework to coding tasks, we analyze the effect of key hyperparameters namely, the decision strategy (majority vote vs. judge decision), the number of debating agents, and the number of debate rounds to identify the optimal configuration (details provided in Section~\ref{method_debate_config}). Using this configuration, we evaluate the applicability and effectiveness of MAD on coding tasks, comparing its performance against SOTA techniques from prior SE studies and single-agent prompting baselines, which naturally motivates our second research question.

\vspace{5pt}
\noindent\textbf{RQ2: What are the debate patterns in MAD that are responsible for the underperforming cases of coding tasks?}

To understand the factors contributing to underperforming cases in MAD for coding tasks, we manually analyzed debate logs using an open-coding approach~\cite{Glaser2016OpenCD} and categorized the debate patterns responsible for the underperforming cases. Beyond categorization, we conducted a root cause analysis of each underperforming debate category, manually examining all 786 underperforming debate logs to identify the underlying causes of each pattern (Table~\ref{tab:root_cause}). To the best of our knowledge, no prior study has analyzed how different debate patterns in MAD impact task performance in SE or examined the root causes of each failure mode of debate patterns. Using these analyses, we aim to refine the MAD to enhance its effectiveness for coding tasks, which leads to our next research question:

% To understand the factors contributing to underperforming cases in MAD for coding tasks, we manually analyzed debate logs using an open-coding approach~\cite{Glaser2016OpenCD} and categorized the debate patterns responsible for the underperforming cases. To the best of our knowledge, no prior study has analyzed how different debate patterns in MAD impact task performance in SE. Using these patterns, we aim to refine the MAD to enhance its effectiveness for coding tasks, which leads to our next research question:

\vspace{5pt}
\noindent\textbf{RQ3: How can we improve the performance of MAD on coding tasks?}

To enhance MAD’s effectiveness on coding tasks, we introduce two targeted strategies that address specific debate patterns observed in RQ2. These enhancements are guided by the judge, who applies task-specific criteria to evaluate the correctness of agents' arguments. 
Our key contributions are:

\begin{itemize}
    \item We investigate the applicability of MAD to coding tasks, extending its application beyond NLP.  
    \item We evaluate MAD on four coding tasks in SE using both proprietary and open-source LLMs.
    \item We analyze underperforming cases in MAD, categorize the debate patterns, and identify the root causes of each debate pattern to understand their impact on task performance.
    \item We propose two enhanced variants of the MAD designed to improve its effectiveness on coding tasks.
\end{itemize}

The remainder of this paper is organized as follows: 
\textbf{Section~\ref{section:related work}} presents previous related studies on MAD. \textbf{Section~\ref{section:methods}} describes the formulation of our MAD variants on four selected coding tasks and our approaches to improve MAD's effectiveness on these tasks. \textbf{Section~\ref{section:setup}} outlines the experimental setup for our study. \textbf{Section~\ref{section:result}} provides the results of our manual analysis and two enhanced variants of MAD. In \textbf{Section~\ref{section:discussion}}, we discuss possible future implications for researchers and practitioners. Finally, \textbf{Section~\ref{section:threats}} presents threats to the validity of our study.

\section{Background \& Related Work}
\label{section:related work}
% Related Work
% 1– sentence summary of each previous research work
% Group them into similar categories/groups
% 1– sentence summary of what they still lack, re-emphasizing and differentiating the value of our work and contribution (however…, In our study…)

\subsection{Multi-Agent Debate}
\label{sec:mad_relatedwork}
MAD systems have gained increasing attention in the NLP literature as they extend beyond LLM-based systems that typically rely on external tool use~\cite{lee2024unified, yang2025agentnet}, the availability of fine-tuning data~\cite{suh2024empirical, murali2024ai, ding2024semcoder}, and prompt design~\cite{wei2022chain, wang2022self}. Specifically, these systems structures debate-style interactions, allowing \llmAgents~to observe, critique, and refine each other's outputs through iterative reasoning and argumentation~\cite{du2023improving, smit2024goingmadlookmultiagent, chan2024chateval}. Prior work on MAD has demonstrated SOTA performance on task such as mathematical reasoning and question answering~\cite{du2023improving, liang-etal-2024-encouraging}, which require diverse reasoning~\cite{smit2024goingmadlookmultiagent, chan2024chateval}, problem-solving~\cite{du2023improving, li2024improving, Xiong_2023}, and decision-making strategies~\cite{liang-etal-2024-encouraging, khan2024debating}. However, existing studies have primarily evaluated MAD in solving general NLP tasks that involve only natural language and arithmetic~\cite{Xiong_2023, estornell2025multi, chan2024chateval, choi2025debate}, overlooking its applicability and effectiveness in coding tasks that involve both natural language and source code. 

% gained attenion in NLP -> we apply to SE -> we improve upon

Designing and developing code during software development typically involve collaboration among developers with diverse expertise~\cite{he2024llm, 7961464}. Developers engage in debate and exchange ideas to reach consensus for tasks such as proposing new features or enhancing existing functionalities~\cite{7961464}. Moreover, many coding tasks in SE are inherently open-ended, allowing multiple valid solutions rather than a single correct answer~\cite{brooks1987no}. Similarly, MAD encourages agents to generate multiple candidate solutions while cross-examining and critiquing one another’s reasoning from diverse perspectives, enabling the exploration of diverse reasoning paths before converging on a solution~\cite{chan2024chateval, du2023improving}. Moreover, structured debate has proven effective for tasks involving open-ended answers, where iterative refinement, argumentative evaluation, and error surfacing can improve final response quality~\cite{liang-etal-2024-encouraging, smit2024goingmadlookmultiagent}. We posit that the automation of coding tasks can benefit from adopting MAD that leverages the collective intelligence and debate-driven synergy of multiple LLM-based agents. However, the effectiveness of MAD on coding tasks has not been extensively investigated.

%new IA
A recent study by Chen et al.~\cite{chen2025debatecoder} evaluated MAD for code generation, conducting debates guided by test case execution. However, the scope of their study was restricted to a single SE task, and it did not investigate the role of debate interactions in shaping MAD’s effectiveness on that task. Analyzing these interactions is crucial because the quality and trajectory of the debate directly determine whether MAD converges on correct solutions or reinforces errors~\cite{khan2024debating, kaesberg-etal-2025-voting, estornell2025multi}. In contrast, our work systematically evaluates MAD across four diverse coding tasks in SE and analyzes how debate dynamics influence its performance, offering broader insights into its applicability in SE.

% Pan et al.~\cite{pan2025multiagent} introduce MAST, a taxonomy of failure modes for Multi-Agent Systems across coding and math problem-solving tasks. While MAST categorizes system-level failures (i.e., Step Repetition and Disobey Task Specification), our work focuses specifically on the MAD interaction pattern, analyzing how agents shift or maintain their positions throughout the debate process. Rather than identifying failure modes at the system level, we identify the root causes underlying each debate interaction pattern that leads to underperformance on coding tasks. Finally, our work goes beyond failure identification by proposing an optimized variant of MAD that directly addresses these root causes to improve performance on coding tasks.

Prior studies in NLP have identified \textit{error propagation} as a key limitation of MAD~\cite{wang2024rethinking, estornell2025multi, Xiong_2023}, where misinformation cascades across agents during debate. This can lead to deviations from an initially correct answer, resulting in an incorrect consensus. Recent studies in NLP~\cite{chen2024magdi, qian2023chatdev, khan2024debating} have proposed various techniques to mitigate error propagation in debate. These techniques primarily focus on modifying agent personality~\cite{liang-etal-2024-encouraging, chan2024chateval, khan2024debating}, communication and information flow~\cite{du2023improving, smit2024goingmadlookmultiagent, Xiong_2023}, or the consensus strategy~\cite{kaesberg-etal-2025-voting, qian2023chatdev, li2023camel, li2024improving, chen2024magdi} of MAD. In contrast, we incorporate debate intervention strategies to prevent suboptimal trajectories, mitigate error propagation, and enable effective interactions among agents. 
% added for comment 3.4
Specifically, we introduce two strategic interventions, namely \textbf{early termination}, which terminates the process once a valid answer is found, and \textbf{random restart}, which re-initiates the process to escape local optima. While these intervention strategies draw inspiration from commonly adopted optimization techniques such as self-refinement~\cite{madaan2023self} and automated prompt optimization~\cite{ahmed2024automatic}, targeted application to the MAD framework to address specific failure modes identified in underperforming debate cases remains unexplored.
To the best of our knowledge, our work is the first to apply debate intervention strategies to improve the outcome of MAD for coding tasks.

\subsection{Selected Coding Tasks}
\label{background_selected_tasks}
We introduce the \textbf{four coding tasks} evaluated in this study and outline the rationale behind their selection.
Two main criteria guided our task selection. First, we selected tasks with characteristics that prior research has shown to be particularly beneficial for MAD. In prior research, MAD has demonstrated effectiveness in mathematical reasoning~\cite{du2023improving, liang-etal-2024-encouraging}, question answering~\cite{du2023improving, liang-etal-2024-encouraging}, iterative reasoning~\cite{du2023improving, li2024improving}, leveraging diverse perspectives~\cite{liang-etal-2024-encouraging, chan2024chateval}, and incorporating critical feedback~\cite{smit2024goingmadlookmultiagent}. Building on these findings, we chose coding tasks that are widely studied in the SE literature~\cite{hou2024large} where recent LLM-based techniques have similarly benefited from iterative reasoning, diverse perspectives, and critical feedback. Second, we emphasized diversity in the tasks to capture different problem-solving capabilities of LLMs. Following a recent survey of LLM applications in SE~\cite{hou2024large}, and based on the two criteria mentioned, we selected two tasks emphasizing reasoning, Code Input Prediction~\cite{ding2024semcoder, jain2025livecodebench} and Code Output Prediction~\cite{ding2024semcoder, chen2024reasoning, gu2024cruxeval}. And two tasks emphasizing understanding, Code Translation~\cite{pan2024lost, yang2024exploring, macedo2024intertrans} and Code Summarization~\cite{ahmed2024automatic, sun2024extractive, virk2025calibration, haque2022semantic}. To the best of our knowledge, this is the first study to investigate the effectiveness of MAD across diverse coding tasks systematically. We aim to understand whether MAD can generalize beyond NLP and provide benefits for coding tasks.

\textbf{Code Input Prediction and Code Output Prediction} involve predicting program behavior without execution~\cite{chen2024reasoning, modelresaon2023pei}. Given a program code, Input Prediction predicts the input that produces a given output, while Output Prediction predicts the output resulting from a given input.
Recent studies have shown that providing task-specific context information (e.g., input–output relationships and input constraints)\cite{ding2024semcoder} and runtime program behavior (e.g., variable values and types at specific statements)\cite{chen2024reasoning} can significantly improve LLM performance on code input and output prediction. While these results are promising, existing techniques are limited to single-iteration, single-model processes and do not explore whether iterative, multi-turn interactions could further enhance LLM reasoning. By contrast, the MAD employs a debate-driven prompting strategy that enables LLMs to iteratively exchange arguments, refine their reasoning, and build a constructive debate history that captures task-specific context~\cite{du2023improving, chen2024magdi, liang-etal-2024-encouraging, chan2024chateval, smit2024goingmadlookmultiagent}. These multi-turn exchanges should hypothetically encourage models to follow a program's execution process more carefully, generate more accurate input and output predictions, and refine those predictions through iterative debate. However, no prior research has investigated this. To address this research gap, in this paper, we investigate the potential of MAD for \textit{Code Input Prediction} and \textit{Code Output Prediction}, evaluating whether its iterative debate process can provide consistent improvements over single-turn, single-model approaches.

% \subsection{Code Translation}
\textbf{Code Translation}, which converts code snippets from one programming language to another~\cite{jiao2023evaluation}, plays a crucial role in modernizing legacy software systems and facilitating seamless cross-platform migration \cite{jiao2023evaluation, yang2024exploring, zhu2022multilingual}. However, manual code translation is both time-consuming and error-prone for developers~\cite{yang2024exploring}. To address this, recent studies have leveraged LLMs to automate code translation, achieving promising results~\cite{pan2024lost, macedo2024intertrans,roziere2021leveraging}. However, studies have shown that LLM-based techniques struggle with various mistakes \cite{pan2024lost,yang2024exploring}, primarily due to their limited understanding of the syntactic and semantic differences among various programming languages, each with distinct grammar and conventions~\cite{pan2024lost, yang2024exploring}.

Recent research has shown promising results for iterative refinement and divide-and-conquer approaches, where tasks are broken down into smaller steps and completed sequentially in LLM-based Code Translation \cite{yang2024exploring, pan2024lost, zhu2024semi}. Notably, MAD allows \llmAgents~to engage in iterative debate, refine each other's responses, and decompose complex tasks into manageable subtasks. This suggests that MAD could improve translation accuracy beyond existing single-LLM approaches. However, no prior work has explored MAD's potential in enhancing Code Translation. Our study addresses this gap by investigating whether the MAD can be effectively applied to automated Code Translation and improve its performance.

% \subsection{Code Summarization}
\textbf{Code Summarization} involves generating natural language descriptions of code functionality~\cite{ahmed2022few, ahmed2024automatic, sun2024source, sun2024extractive}. 
Source code comments written in natural language aid program comprehension and significantly enhance software maintainability~\cite{haiduc2010supporting, kilic2024source, linares2015developers}. However, developers often lack the time or motivation to write comments \cite{kilic2024source, sourcecodeFakhoury}, and as projects evolve, existing comments can become outdated \cite{briand2003software}. To address these challenges, researchers have developed Code Summarization techniques that automatically generate natural language descriptions of code functionality \cite{ahmed2022few, ahmed2024automatic, sun2024source, sun2024extractive}. Given that LLMs are trained on vast natural language and code corpora, they achieve SOTA performance through prompting without updating model weights~\cite{sun2024source, ahmed2024automatic}. 

However, current LLM-based code summarization techniques typically rely on a single model. While iterative refinement of LLM-generated summaries has shown promise \cite{li2021editsum, ahmed2024automatic, sun2024extractive}, the potential of multiple \llmAgents~collaborating and engaging in iterative debate for Code Summarization remains unexplored. MAD offers a structured approach that integrates diverse arguments and knowledge across multiple agents, potentially enhancing summarization quality. In this work, we conduct an empirical analysis to investigate whether MAD improves Code Summarization performance.

% \section{Experimental Setup}
% \label{section:setup}
% \input{section/03setup}

\section{Methodology}
\label{section:methods}
% Methodology
% Data collection - provide reasoning for choice of selected projects
% cite papers that use a similar dataset or similar analysis pattern
% Table summarizing the project statistics

In this work, we aim to assess the effectiveness of MAD in four coding tasks: Code Input Prediction, Code Output Prediction, Code Translation, and Code Summarization. We begin by describing the formulation of the MAD, adapted from NLP research (RQ1). We then analyze debate interaction histories to identify patterns contributing to task under-performance (RQ2), and propose improvements to MAD based on these patterns (RQ3).

% \vspace{5pt}
\subsection{Methodology to Answer RQ1}
\label{method_debate_stage}

We adapt the MAD framework from NLP research~\cite{du2023improving, liang-etal-2024-encouraging}, structuring debates into multiple stages aligned with SE-specific subtasks. At each stage, agents engage in focused debates to iteratively exchange information and refine their contributions toward the overall task. Following NLP best practices~\cite{xue2024decompose, wei2022chain}, each agent's output at every stage consists of three components: (1) Task Output, the agent's response to the subtask; (2) Position, indicating agreement or disagreement with the responses from agents who already give responses in the debate (i.e., \textit{``I agree with the preceding agent's argument''} or \textit{``I disagree with Agents 2's perspectives because...''}) and (3) Explanation, providing the rationale for its stance and response.

\subsubsection{MAD Implementation}\label{method_debate_prompt}

Following~\cite{du2023improving, liang-etal-2024-encouraging}, our MAD implementation consists of three debate agents and one judge agent. The debate agents observe, critique, and refine each other's responses iteratively. The judge agent determines which agent presents the acceptable response based on the task objectives and debate topic. This response is then sent to the following debate stage (Section \ref{sec:decompos}) for further refinement. To detail, we constructed unique prompts for each debate and judge agent and prompted the LLM to simulate a Multi-Agent Debate. For debate agents, we provided role descriptions, task instructions, the code snippet, the expected output format, and the goal of the debate stage (Section~\ref{sec:decompos}) in the prompts, following the prompting practices proposed by~\cite{liang-etal-2024-encouraging} to facilitate agent debate.
For the judge agent, we adopted the prompting structure proposed by~\cite{smit2024goingmadlookmultiagent}, which includes a role description and a list of 
key assessment factors to determine the best response. These factors are task-specific and were chosen based on well-established evaluation criteria from prior work that studied quality dimensions for the selected tasks. Table~\ref{tab:judge_criteria} lists these factors.

\captionsetup{skip=1pt}
\begin{table}[t]
\setlength{\tabcolsep}{1pt}
    \centering
    \caption{
    \centering
    Judge Assessment Factor Definitions for Code Input Prediction (CIP), \\ Code Output Prediction (COP), Code Summarization (CS), and Code Translation (CT)}
    \resizebox{0.75\linewidth}{!}{
    \begin{tabular}{p{0.9cm} p{3.5cm} p{7cm}}
        \toprule
        \textbf{Task} & \textbf{Assessment Factor} & \textbf{Definition} \\
        \midrule
        \multirow{2}{*}{\textbf{CIP}} 
        & Input Correctness & Correctness of the input based on code logic. \\
        & Compatibility & Adherence to the code’s input constraints. \\
        \midrule
        \multirow{2}{*}{\textbf{COP}} 
        & Output Correctness & Correctness of the output based on code logic. \\
        & Coherence & Consistency with the intended code behavior. \\
        \midrule
        \multirow{2}{*}{\textbf{CT}} 
        & Translation Accuracy & Consistency in syntax and structure. \\
        & Functional Correctness & Maintaining the code functionality. \\
        \midrule
        \multirow{3}{*}{\textbf{CS}} 
        & Expressiveness & Clarity and readability of summary. \\
        & Content Adequacy & Coverage of key class details. \\
        & Conciseness & Avoidance of unnecessary details. \\
        \bottomrule
    \end{tabular}
    }
    \label{tab:judge_criteria}
\end{table}

For Code Input Prediction, the task-specific assessment factors include \textit{correctness} and \textit{compatibility}. For Code Output Prediction, we assess \textit{correctness} and \textit{coherence}. These dimensions are widely used to evaluate the understanding of input constraints, execution of internal logic, and reasoning over the input-output behavior of a program~\cite{ding2024semcoder, gu2024cruxeval}.
For Code Translation, we focused on \textit{translation accuracy} and \textit{functional correctness}, as recommended in recent studies that identify these dimensions as essential for preserving the intended behavior of the translated code~\cite{jiao2023evaluation}.
For Code Summarization, we employed \textit{expressiveness}, \textit{content adequacy}, and \textit{conciseness}, which are widely used in evaluating summary quality and have been shown to align well with human preferences~\cite{hu2022correlating}.

% It is worth noting that \textit{Position} refers to whether it agrees with the first agent's \textit{Task Output} or not 

At each stage, the first agent initiates the debate by generating the \textit{Task Output} and \textit{Explanation}. The second agent then observes the first agent's response and produces a \textit{Position} indicating whether it agrees or disagrees with the first agent, and provides its own \textit{Task Output} and an \textit{Explanation} for its response. Similarly, the third agent considers both preceding responses, generates a \textit{Position} to indicate its stance with respect to the two agents, and produces its own \textit{Task Output} and \textit{Explanation}. At the end of the stage, the judge agent evaluates all responses and determines an acceptable answer based on task-specific assessment factors (Table~\ref{tab:judge_criteria}). The judge produces a \textit{Winning Agent}, the agent who provided the acceptable response, and an \textit{Explanation} for selecting it as the most acceptable answer. The judge’s response, including the \textit{Winning Agent}, the \textit{Explanation}, and the \textit{Task Output} from the winning agent, is then provided to the first agent in the next debate stage, repeating the same procedure. This serves as the initial context for the following debate stages, enabling agents to iteratively refine and build upon the previous debate outcomes.

% The second agent then observes the first agent's response and produces a \textit{Task Output}, \textit{Position}, and \textit{Explanation}. 
% Specifically, the \textit{Position} of the second agent refers to whether it agrees with the first agent's \textit{Task Output} and \textit{Explanation}, and it produces its own \textit{Task Output} and \textit{Explanation}. 

% \begin{figure}[!ht]
% \centering
% \includegraphics[width=0.6\linewidth]{figures/methods/debate_prompt.pdf}
% \includegraphics[width=0.6\linewidth]{figures/methods/judge_prompt.pdf}
% \caption{Overview of prompt used for \llmAgents~}
% \label{fig:method_prompt}
% \end{figure}

\captionsetup{skip=0pt}
\begin{figure}

\lstset{
  language=Python,
  basicstyle=\ttfamily\scriptsize,
  keywordstyle=,
  showstringspaces=false,
  columns=fullflexible,
  escapeinside={(*@}{@*)},
}

\begin{tcolorbox}[
    enhanced, 
    colback=gray!5, 
    colframe=cyan!50!black, 
    colbacktitle=cyan!70!black, 
    coltitle=white, 
    title style={font=\bfseries},
    boxsep=0pt,        
    % left=2pt,         
    right=0pt,         
    % top=2pt,          
    bottom=0pt,        
    arc=4pt    
]

\begin{minipage}[t]{0.40\linewidth}
\textbf{Debater System Prompt:}
\begin{lstlisting}
You are a software developer participating 
as a debater. There are (*@\textcolor{blue}{\{num\_agents\}}@*) debaters 
involved in a (*@\textcolor{blue}{\{task\}}@*) debate challenge. 
You are given a code file in (*@\textcolor{blue}{\{code\_language\}}@*) 
to perform (*@\textcolor{blue}{\{task\}}@*), which will be conducted 
in debate format.
\end{lstlisting}

% \vspace{0.1em}
\textbf{Agent 1 User Prompt:}
\begin{lstlisting}
You are an expert in (*@\textcolor{blue}{\{role\_assignment\}}@*)
Your task is to (*@\textcolor{blue}{\{subtask\_specification\}}@*)
Original code: (*@\textcolor{blue}{\{code\_snippet\}}@*)
# instruction for response and ouput format
\end{lstlisting}

% \vspace{0.1em}
\textbf{Agent 2 and 3 User Prompt:}
\begin{lstlisting}
You are an expert in (*@\textcolor{blue}{\{role\_assignment\}}@*)
Your task is to (*@\textcolor{blue}{\{subtask\_specification\}}@*)
Present your explanation or counter-argument 
to the preceding agents' responses.
Debate History: (*@\textcolor{blue}{\{previous\_agent\_response\}}@*)
Original code: (*@\textcolor{blue}{\{code\_snippet\}}@*)
# instruction for response and ouput format
\end{lstlisting}

% % \vspace{0.1em}
% \textbf{Agent 3 User Prompt:}
% \begin{lstlisting}
% You are an expert in (*@\textcolor{blue}{\{role\_assignment\}}@*)
% Your task is to (*@\textcolor{blue}{\{subtask\_specification\}}@*)
% Present your explanation or counter-argument 
% to the preceding agents' responses.
% Debate History: (*@\textcolor{blue}{\{previous\_agents\_response\}}@*)
% Original code: (*@\textcolor{blue}{\{code\_snippet\}}@*)
% # instruction for response and ouput format
% \end{lstlisting}

\end{minipage}
\hfill
\begin{minipage}[t]{0.45\linewidth}
\textbf{Judge System Prompt:}
\begin{lstlisting}
You are a moderator.
There are (*@\textcolor{blue}{\{num\_agents\}}@*) debaters involved 
in (*@\textcolor{blue}{\{task\}}@*) debate challenge.
Debaters will present arguments on 
(*@\textcolor{blue}{\{overall\_task\_description\}}@*)
# Explain the goal of the overall task goal
\end{lstlisting}

% \vspace{0.1em}
\textbf{Judge User Prompt:}
\begin{lstlisting}
Original code: (*@\textcolor{blue}{\{code\_snippet\}}@*)
Debate history: (*@\textcolor{blue}{\{agent\_debate\_history\}}@*)
Evaluate based on (*@\textcolor{blue}{\{task\_criteria\}}@*)
# instruction for response and ouput format
\end{lstlisting}
\end{minipage}

\end{tcolorbox}

\caption{
\centering
System and User prompt templates for Debate and the Judge Agents. \\ \textcolor{blue}{Blue \{text\}} representing variables filled in during prompt construction.}
\label{fig:debate_prompt}
\end{figure}

Figure~\ref{fig:debate_prompt} displays the prompts used for the debate and judge agents in our experiment. These prompts were based on established NLP prompting strategies~\cite{du2023improving, liang-etal-2024-encouraging} to facilitate effective debate among the \llmAgents. We did not conduct an ablation study on individual prompt elements, as each component of the prompts was carefully designed to ensure proper role adherence and interaction within the MAD and was considered essential for realizing the full MAD process.

\subsubsection{MAD Configurations}
\label{method_debate_config}

Following prior NLP research~\cite{du2023improving, liang-etal-2024-encouraging}, the configurations of MAD include the decision strategy (i.e., majority vote or judge decision), the number of agents participating in the debate, the temperature value of the LLMs, and the number of debate rounds (i.e., the number of debate iterations to solve a single data sample). We analyzed the impact of these hyper-parameters on four coding tasks to identify the configuration that yields optimal performance. Specifically, we first collected configurations from prior MAD research~\cite{qian2023chatdev, du2023improving, liang-etal-2024-encouraging, li2024improving} to define a search space. The search space includes two decision strategies (majority vote and judge decision), \textit{(2, 3, 4)} for the number of debate agents, \textit{(0, 0.2, 0.7)} for the temperature of the LLMs, which determines the variability and randomness in the output, and \textit{(2, 3, 4, 5)} for the number of debate rounds.

\captionsetup{skip=5pt}
\begin{figure}[!ht]
\centering
\includegraphics[width=0.8\linewidth]{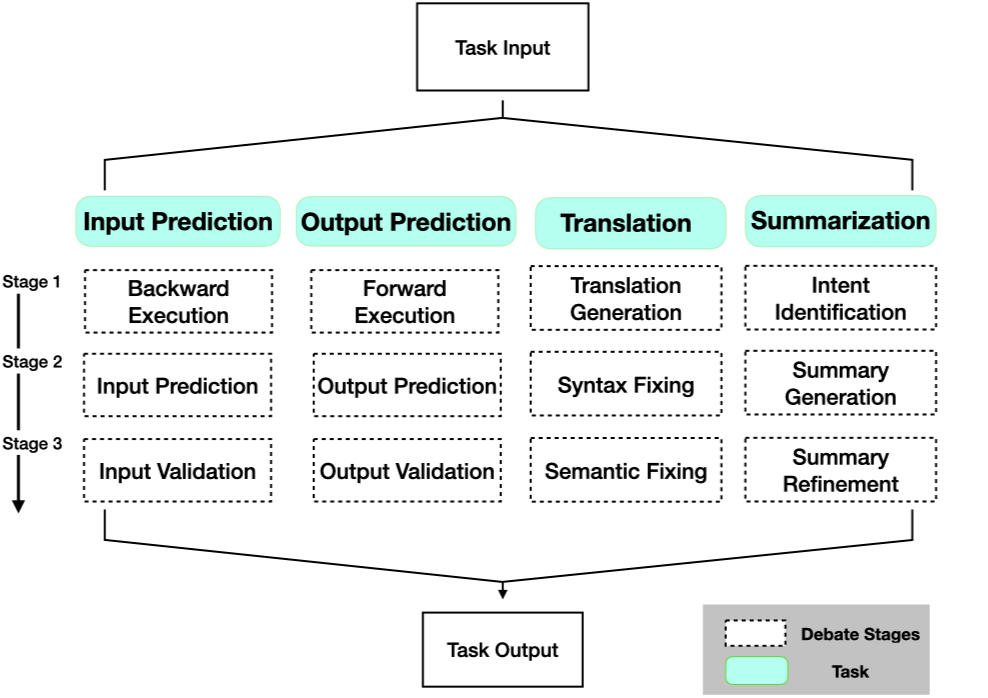}
\caption{Overview of Debate Stages in MAD}
\label{fig:method_mad}
\end{figure}

\subsubsection{Task Decomposition} \label{sec:decompos}

% structuring debates into multiple stages aligned with SE-specific subtasks.
We decomposed each of the four coding tasks into three subtasks, following
prior studies~\cite{ding2024semcoder, chen2024reasoning, pan2024lost, sun2024source} that outline the procedures of automated techniques. To detail, Code Input Prediction consist of three subtasks~\cite{ding2024semcoder}: (1) \textbf{Backward Execution Reasoning} performs backward step-by-step execution from the output to the initial state of the program, outlining the backward tracing of program (2) \textbf{Input Prediction} predicts the input values given the output of program execution and (3) \textbf{Input Validation} refines the predicted input values to verify whether they satisfy the constraints of the program.

Similarly, Code Output Prediction consists of three subtasks~\cite{ding2024semcoder}: (1) \textbf{Forward Execution Reasoning} performs forward step-by-step execution of the program to outline its operational logic (i.e., line of code being executed) (2) \textbf{Output Prediction} predicts the execution output values given the input of the program and (3) \textbf{Output Validation} refines the predicted output values to verify whether they align with the intended functionality of the program. 

The three subtasks of Code Translation include~\cite{pan2024lost}: (1) \textbf{Translation} converts the original code into the target programming language, (2) \textbf{Syntactic Error Resolution} resolves syntactic errors (i.e., missing semicolons or incorrect indentation) in the translated code, and (3) \textbf{Semantic Error Resolution} resolves semantic errors (i.e., wrong variable or operator~\cite{pan2024lost}) in the translated code. 

Lastly, we decomposed Code Summarization into three subtasks~\cite{sun2024source}: (1) \textbf{Intent Identification} identifies the perspective from which the code should be summarized. Specifically, we allowed the debate agents to debate on the intent categories widely adopted by Code Summarization studies \cite{sun2024source, mu2023developer}:  ``What'' (description of the functionality), ``Why'' (the design rationale of the code), ``How-to-use'' (description of the usage), ``How-it-is-done'' (implementation details of the functionality), ``Property'' ( the properties of the code including pre-conditions or post-conditions) and ``Others'', (2) \textbf{Summary Generation} generates a code summary from the identified intent, and (3) \textbf{Summary Refinement} evaluates and refines the generated summary to improve its semantic relevance to the code. Consequently, \llmAgents~ debate with each other to solve subtasks, with the judge agent deciding on the acceptable response to move on to the next subtask. Hence, each debate stage focuses on solving a single subtask, and the subtasks are solved in a sequential order. Figure~\ref{fig:method_mad} displays an overview of MAD for our selected four coding tasks.

%\hl{At each stage, the debates can be multiple iterative rounds?}.

%\hl{match the order of the picture and the order of the text, left to right}

%\hl{CITE needed for semantic errors}

\vspace{5pt}
\subsection{Methodology to Answer RQ2}
\label{method_debate_cateogries}

To answer RQ2, our analysis followed three steps: (1) manually categorizing agent interaction patterns in underperforming debate logs, (2) scaling the categorization to all underperforming cases using LLM-based annotation, and (3) identifying the root causes underlying each failure category.

% step 1
\subsubsection{Manual Categorization}\label{method:manual_categorization}We manually analyzed the debate logs of underperforming cases, where the judge agents determined that none of the agents' \textit{Task Output} was acceptable based on judge assessment factors in Table \ref{tab:judge_criteria} (Section \ref{method_debate_prompt}). We defined underperforming cases as instances where the judge found no acceptable answer from the MAD agent debate. The total number of underperforming cases was 786, with 100 for the Code Input Prediction task, 176 for the Code Output Prediction task, 431 for the Code Translation Task, and 79 for the Code Summarization task. We then manually categorized the agent interaction patterns during the debates. The goal of this manual analysis step was to identify the patterns most responsible for the underperforming cases.

Based on the agent interaction logs produced during the debates in RQ1, two authors, each with over five years of programming experience, independently analyzed the logs to identify categories of interaction patterns following an open coding protocol~\cite{Glaser2016OpenCD}. To ensure the manual analysis remained manageable while retaining statistical validity, we randomly sampled 260 debate logs from the total of 786 underperforming cases (95\% confidence level with a 5\% margin of error). Following independent coding, the two authors engaged in multiple rounds of discussion to resolve discrepancies through continuous comparison and negotiated agreement~\cite{Forman2007QualitativeCA}, iteratively refining the categories. This process was repeated until a full consensus was reached on the final set of interaction pattern categories, which represent the trajectories by which agents shift or maintain their points of view throughout the debate. These categories are presented in Section~\ref{section:result}.

% step 2
\subsubsection {LLM-based Annotation.}
\label{sec:method_llm_annotation}
Equipped with the finalized categories, our next goal was to analyze all debate interactions. To scale the annotation process beyond manual labeling, we adopted an LLM-based annotation method to categorize the remaining underperforming debate logs. Recent research has shown that LLMs can effectively reduce the human effort required for manual dataset annotation while maintaining high-quality results~\cite{ahmed2024can, zhou2025}.

Following best practices for LLM-based annotation of SE artifacts~\cite{ahmed2024can}\cite{zhou2025}, we designed a detailed prompt that included (1) a description of the SE task being debated, (2) the format of the debate logs, (3) definitions of the debate interaction categories, (4) the expected output format (i.e., the identified categories of debate interaction patterns), and (5) four few-shot examples of debate logs paired with their assigned categories, which we curated from our initial human-coded set.

To select the base model for performing the LLM-based annotation, we experimented with three models that demonstrated SOTA performance in annotating SE artifacts, namely GPT-4o~\cite{gpt-4o}, Gemini~\cite{team2023gemini}, and DeepSeek~\cite{deepseek_chat}, as reported in prior studies~\cite{ahmed2024can, zhou2025}.
We validated the annotation performance of each model on a subset of 260 debate logs that had been labeled by human annotators and selected the model with the highest agreement with human annotations to apply to the full set of underperforming cases. Underperforming cases are defined as instances where the judge found no acceptable answer from the MAD agent debate, yielding no task outcome (details on the underperforming debate cases are provided in Section~\ref{method_debate_cateogries}).

% added for reviewer comment 3.6
As a result, GPT-4o achieved an inter-rater reliability (IRR) score of 87.3\%, Gemini achieved an IRR score of 83.8\%, and DeepSeek achieved an IRR score of 69.2\%. Following Ahmed et al.~\cite{ahmed2024can}, an agreement score above 0.5 indicates that LLM-provided annotations can safely replace human ratings for SE artifacts. These high IRR scores demonstrate that all three models can serve as reliable annotators for our dataset. We employed GPT-4o to categorize the remaining debate logs, as it achieved the strongest agreement with human judgments.

In total, we produced categorizations for 786 underperforming debate cases. The final results of this categorization are presented in Section~\ref{section:result}. We exclusively used LLM-based annotation to label debate logs using the finalized categories derived from the human annotation process. Hence, no additional categories were identified by the LLMs during the annotation process. The prompt used to conduct LLM-based data annotation for categorizing the underperforming cases of debate is shown in Figure~\ref{fig:annotation_prompt}.

\captionsetup{skip=0pt}
\begin{figure}

\lstset{
  language=Python,
  basicstyle=\ttfamily\scriptsize,
  keywordstyle=,
  showstringspaces=false,
  columns=fullflexible,
  escapeinside={(*@}{@*)},
}

\begin{tcolorbox}[
    enhanced, 
    colback=gray!5, 
    colframe=cyan!50!black, 
    % title=Data Annotation Prompt Template (GPT-4o),
    colbacktitle=cyan!70!black, 
    coltitle=white, 
    title style={font=\bfseries}
    boxsep=0pt,        
    % left=2pt,         
    right=0pt,         
    top=1pt,          
    bottom=0pt,        
    arc=4pt,    
    ]

\textbf{System Prompt:}
\begin{lstlisting}
You are provided a history of debate between three agents on the topic of (*@\textcolor{blue}{\{task\}}@*) task.
The debate log displays responses by three agents showing their task output, position, and explanation.
# include description of task output, position, and explanation.
\end{lstlisting}

\vspace{0.4em}
\textbf{User Prompt:}
\begin{lstlisting}
Categories:(*@\textcolor{blue}{\{debate\_pattern\_categories\}}@*) # include definition of each categories
Few-shot Examples:(*@\textcolor{blue}{\{few\_shot\_examples\}}@*)
Debate History:(*@\textcolor{blue}{\{full\_debate\_logs\}}@*)
Your response must return exactly one category from the list above.
 # instruction for response and ouput format
\end{lstlisting}
\end{tcolorbox}

\caption{
\centering
Overview of the prompt used for GPT-4o data annotation of debate interaction patterns.\\ \textcolor{blue}{Blue \{text\}} representing variables filled in during prompt construction.}
\label{fig:annotation_prompt}
\end{figure}

% code complexity measure, reviewer comment 3.6
\subsubsection{Complexity metrics}\label{complexity_metrics}
To investigate whether the identified underperforming patterns (Section~\hyperlink{target_debate_categories}{5, RQ2 Result}) stem from inherent code complexity or from agent interactions in the MAD process, we analyze the complexity of code samples by comparing those used in underperforming debates with those used in successful (i.e., debate with a valid task outcomes) debates. Following Kasto et al.~\cite{kasto2013measuring}, we adopt five metrics: \textbf{Number of Statements}, \textbf{Number of Operands}, \textbf{Cyclomatic Complexity (CC)}, \textbf{Average Nested Block Depth}, and \textbf{Average Number of Parameters}. These metrics are used to compute the complexity score of each code sample in our MAD experiments (Table~\ref{tab:task_summary}). We then compare the mean complexity scores between successful and underperforming debates.

% step 3
\subsubsection{Root Cause Analysis.} 

\label{root_cause_methods} Following the LLM-based categorization, we conducted a deeper analysis to identify the root causes underlying each failing debate category (namely, forceful agreement, ending divergence, and prolonged disagreement). Three authors, each with over five years of programming experience, independently analyzed underperforming debate logs for each debate category following an open coding protocol~\cite{Glaser2016OpenCD}. Specifically, the authors analyzed debate logs in increments of 10 selected at random for each debate category, discussing and resolving discrepancies through continuous comparison and negotiated agreement~\cite{Forman2007QualitativeCA} until no new root causes emerged. Once full consensus was reached on the final set of root causes, the remaining debate logs within each category were split equally among the three authors for full annotation. This process was repeated for each failing debate category for a total of 786 underperforming debate logs (details on the definition of underperforming debates are provided in Section~\ref{method:manual_categorization}). The results of this root cause analysis are summarized in Table~\ref{tab:root_cause}, and the definitions of each root cause are provided in Table~\ref{tab:root_cause_def}.

% \label{root_cause_methods}{Following the LLM-based categorization, we conducted a deeper analysis to identify the root causes underlying each failing debate category (namely, forceful agreement, ending divergence, prolonged disagreement). Three authors, each with over five years of programming experience, independently analyzed underperforming debate logs for each failure category following an open coding protocol~\cite{Glaser2016OpenCD}. Specifically, the authors analyzed debate logs in increments of 10 selected at random for each debate category, discussing and resolving discrepancies through continuous comparison and negotiated agreement~\cite{Forman2007QualitativeCA} until no new root causes emerged. Once a full consensus was reached on the final set of root causes, the remaining debate logs within each category were split equally among the three authors for full annotation. This process was repeated for each debate category (namely, forceful agreement, ending divergence, prolonged disagreement) for the total of underperforming debate logs set up 786 debate logs (details on the definition of underperforming debate in section \ref{method:manual_categorization}. The results of this root cause analysis are summarized in Table~\ref{tab:root_cause} and the definition of each root causes are written in Table~\ref{tab:root_cause_def}.

\subsection{Methodology to Answer RQ3}
\label{sec:method_rq3}

This section presents mitigation strategies targeting the identified debate categories of underperforming cases in MAD, as shown in Section~\ref{section:result}. Guided by the engineering optimization strategies such as early termination and feedback-based correction mechanisms, we introduce two enhanced MAD variants to prevent agents from error propagation and converging on incorrect conclusions. Furthermore, to gain an understanding of token usage and API calls (i.e., inference calls to query the LLM), we calculate the total number of tokens used and the total number of API calls made for running our experiments, which are reported in  Table~\ref{tab:mad_token_usage}.

\textbf{Early Termination:} In the \textit{Ending Divergence} (Section~\ref{section:result}) debate category, we observed that although agents initially agree on an acceptable or partially correct response, they tend to diverge as the debate progresses. Instead of refining or improving the initial answer, the agents introduce increasingly different and often less accurate responses over successive debate rounds. This divergence can result in underperforming cases, especially when the final, diverged response is selected as the outcome of the debate. 

Similarly, in the \textit{Forceful Agreement} category, agents start from a point of disagreement, which can be productive for exploring different perspectives. However, rather than debating toward a correct solution, agents may converge on an incorrect response, reinforcing each other's flawed reasoning and resulting in underperformance. Thus, both categories illustrate how breakdowns in debate dynamics, whether through unnecessary divergence or misguided agreement, can negatively impact the quality of outcomes produced by the MAD.

To mitigate these underperforming cases of debate, we introduce a MAD variant equipped with an optimization mechanism that terminates the debate as soon as the judge identifies an acceptable answer. Specifically, to prevent unnecessary divergence and misguided agreement, the debate is terminated once the judge identifies an acceptable answer based on key assessment factors (Section~\ref{method_debate_prompt}). This ensures that other agents do not further influence the response, preventing unnecessary debate rounds.

\textbf{Extended Reflection:} In the \textit{Prolonged Disagreement} debate category (Section~\ref{section:result}), we observed that agents typically start in disagreement and persist in this state without reaching any consensus by the end of the debate. This category is often characterized by agents engaging in argumentative exchanges, resisting each other’s reasoning, and deviating from the goal of collaboratively refining their responses. Rather than working toward a stronger, unified answer, agents remain entrenched in their initial positions, limiting the debate’s capacity to generate meaningful improvements. Such behavior also highlights a lack of collaborative synergy between the agents, ultimately undermining the effectiveness of the MAD in these cases.

To address this issue and reduce agent stubbornness that can negatively impact debate outcomes, we introduced a \textit{``no winner''} option in the judge prompt when none of the agents' responses are acceptable. When invoked, the judge is further prompted to provide constructive feedback based on the assessment factors (Table~\ref{tab:judge_criteria}) to help agents reflect on their responses and understand the shortcomings. This optimization mechanism includes a \textit{``retry opportunity''} for the debate agents, allowing them to re-engage in the same subtask while incorporating the judge's feedback to revise and improve their responses for \textit{``feedback-based correction```} in MAD. This approach aims to guide the debate back on track and foster more collaborative and productive interactions, reducing instances where agents fixate on initial incorrect answers. 

\begin{algorithm}
\caption{MAD Enhancement: Extended Reflection}
\label{method_approach3}
\begin{small} % Start small here
\begin{algorithmic}[1]
\Require Task $t$, Number of Debaters $N$, Maximum Retry $R$
\Procedure{Extended\_MAD}{$t, N, R$}
    \State $J$ \Comment{Initialize the judge}
    \State $D \gets [D_1, \dots, D_N]$ \Comment{Initialize debaters}
    \State $Stages \gets GetStage(t)$ \Comment{Initialize stages for task}

    \For{$S_i$ in $Stages$}
        \State $r \gets 0$ \Comment{Retry count for each stage}
        \State $H \gets [t]$ \Comment{Initialize debate history}

        \While{$r < R$}
            \State $H \gets \text{Debate}(D, S_i, H)$ \Comment{Feedback provided}
            \State $winner, feedback_J \gets J_d(H)$ \Comment{Judge response}

            \If{winner}
                \State \textbf{return} GetAnswer(winner) \Comment{End debate}
            \EndIf
            
            \State $H \gets feedback_J$ \Comment{Append judge feedback}
            \State $r \gets r + 1$
        \EndWhile
    \EndFor
\EndProcedure
\end{algorithmic}
\end{small} % End small here
\end{algorithm}

Algorithm~\ref{method_approach3} illustrates the detailed process of our improved MAD variant by \textit{Extended Reflection}. Inspired by NLP prompting~\cite{madaan2023self}, which leverages the self-refining ability of LLMs, \textit{Extended Reflection} is introduced to encourage iterative refinement between agents while reducing their stubbornness, which may negatively impact the debate outcome. During the debate process, if the judge determines that no acceptable answer is present after evaluating all of the agents' responses (line 9-10), the debate restarts with feedback from the judge on why the prior responses were incorrect (line 13-14). In this MAD variants, the judge provides feedback based on the assessment factors (Table~\ref{tab:judge_criteria}) to help debate agents reflect on their responses and refine them. This approach offers a \textit{``retry opportunity''}(line 8), enabling debate agents to restart the debate while incorporating feedback from the judge (line 13). This mechanism prevents agents from remaining deep-rooted in their initial positions and encourages collaborative synergy by prompting them to learn from their mistakes and the judge’s feedback.

\section{Experimental Setup}
\label{section:setup}
\subsection{Baselines} \label{sec:baselines}

% \hl{TODO: add rationale for including the cr task, tone should be two task to begin with dont put them as subtasks of a task, add to reason pick two from understanding and pick two from generation category. make it bold that SOTA will not be about JAVA out of scope of this work}

\captionsetup{skip=1pt}
\begin{table*}[b]
\centering
\caption{Summary of Tasks with Datasets, Languages, and Models used}
\label{tab:task_summary}
\resizebox{\linewidth}{!}{
\setlength{\tabcolsep}{2pt}
\begin{tabular}{lccccccc}
    \toprule
    \textbf{Task} & \textbf{Dataset} & \textbf{\# Samples} & \textbf{Languages} & \textbf{Subject LLMs} & \textbf{SA Baseline} & \textbf{SE Baseline}\\
    \midrule
    \textbf{CIP} & CRUXEval~\cite{gu2024cruxeval} & 800 & Python & DeepSeek-Coder-6.7B\cite{deepseek6_7}, GPT-4o\cite{gpt-4o} & CoT~\cite{wei2022chain}, SC~\cite{wang2022self} & SemCoder~\cite{ding2024semcoder} \\
    \midrule
    \multirow{2}{*}{\textbf{COP}} & CRUXEval~\cite{gu2024cruxeval} & 800 & Python & \multirow{2}{*}{DeepSeek-Coder-6.7B\cite{deepseek6_7}, GPT-4o\cite{gpt-4o}} & \multirow{2}{*}{CoT~\cite{wei2022chain}, SC~\cite{wang2022self}} & SemCoder~\cite{ding2024semcoder} \\
    & LiveCodeBench~\cite{gu2024cruxeval} & 479 & Python & & & \\
    \midrule
    \multirow{3}{*}{\textbf{CT}} & AVATAR~\cite{ahmad2023avatarparallelcorpusjavapython} & 250, 250 & Python, Java & \multirow{3}{*}{Qwen2.5-Coder-32B\cite{hui2024qwen2}, GPT-4o\cite{gpt-4o}} & \multirow{3}{*}{CoT~\cite{wei2022chain}, SC~\cite{wang2022self}} & \multirow{3}{*}{PLTranslation~\cite{pan2024lost}} \\
    & CodeNet~\cite{puri2021codenet} & 200, 200 & Python, Java & & \\
    & EvalPlus~\cite{evalplus2023} & 164, 164 & Python, Java & & \\
    \midrule
    \textbf{CS} & CodeSearchNet~\cite{husain2020codesearchnetchallengeevaluatingstate} & 665, 665 & Python, Java & CodeLlama-7B\cite{CodeLlama7bInstruct}, GPT-4o\cite{gpt-4o} & CoT~\cite{wei2022chain}, SC~\cite{wang2022self} & ASCS~\cite{sun2024source} \\
    \bottomrule
\end{tabular}
}
\end{table*}

To evaluate the performance of MAD on four coding tasks, we compare against two categories of single-agent baselines: (1) SOTA prompting techniques, namely Chain-of-Thought~\cite{wei2022chain} and Self-Consistency~\cite{wang2022self}, and (2) SOTA Single LLM-based SE methods~\cite{pan2024lost, sun2024source, ding2024semcoder}, ensuring a fair comparison by using the same datasets and models as those adopted in prior work. In our literature review, we found no prior work that provides established multi-agent baselines for the coding tasks included in our study. As a result, we adopt the best-performing single-agent LLM-based methods from prompting technique and SE literature as SOTA baselines.

For Code Input Prediction and Code Output Prediction, we focused exclusively on the Python language, as the SOTA baseline technique is designed for Python datasets, and extending it to other languages is beyond the scope of this work. 
For Code Translation and Code Summarization, we included Python and Java in our experiments, for their widespread use in software development~\cite{ieeeProgrammingLanguages, tiobeTIOBEIndex}. We limited our study to these two languages due to the high computational cost of running MAD~\cite{faiz2023llmcarbon, li2024agentsneed}.

For our experimental dataset, we evaluated MAD on all datasets used in the SOTA SE baselines~\cite{pan2024lost, sun2024source, ding2024semcoder} that provided systematic evaluation of the result.
For Code Input Prediction, we used the CRUXEval~\cite{gu2024cruxeval} with a total of 800 samples. For Code Output Prediction, we used the CRUXEval~\cite{gu2024cruxeval} and LiveCodeBench~\cite{jain2025livecodebench} datasets, with a total of 1,279 samples. 
For Code Translation, we used the AVATAR, CodeNet, and EvalPlus datasets, resulting in a total of 1,228 samples. 
For Code Summarization, we used the CodeSearchNet dataset~\cite{husain2020codesearchnetchallengeevaluatingstate} and randomly selected 1,330 samples for Python and Java (99\% confidence level with a 5\% margin of error).
Table~\ref{tab:task_summary} summarizes the baselines and datasets used in our study.

\subsection{Experimental LLMs}

Open-source LLMs have demonstrated performance on par with proprietary models like GPT-4~\cite{gpt-4o} for coding tasks, offering a cost-effective alternative for building LLM-based systems that require multiple inferences with long-context prompts~\cite{imani2024context, faiz2023llmcarbon}. To ensure diversity in model selection, we chose one proprietary and one open-source LLM for our experiment. We selected the best-performing models based on their strong instruction-following capabilities, support for large context windows, and SOTA performance in both code-related tasks and natural language understanding. Table~\ref{tab:task_summary} outlines the LLM models used for the four selected coding tasks in our study.

For the Code Input and Output Prediction task, we selected DeepSeek-Coder Instruct 6.7B\cite{deepseek6_7} as the open-source model, consistent with the SOTA baseline\cite{ding2024semcoder}, to ensure a fair comparison.
%For Code Input and Output Prediction, we selected DeepSeek-Coder Instruct 6.7B~\cite{deepseek6_7} for open-source model, which is the same model used in the SOTA baseline~\cite{ding2024semcoder}, to ensure a fair comparison. 
For the proprietary model, we replicated SOTA study using GPT-4o~\cite{gpt-4o}, which is widely adopted in the literature~\cite{pan2024lost, ahmed2024can, liang-etal-2024-encouraging} and has demonstrated strong capabilities for code-related tasks~\cite{xia2024agentless, chen2024reasoning, pan2024lost, ahmed2024can, chen2025debatecoder, sun2024source}.
For Code Translation, we selected GPT-4o~\cite{gpt-4o}, the best-performing proprietary LLM used in the SOTA technique proposed by Pan et al.~\cite{pan2024lost}. However, the open-source models used in the SOTA Code Translation baseline~\cite{pan2024lost} lacked sufficient context window capacity to support iterative debate (i.e., they could not sustain multiple debate rounds without exceeding context limits). Since MAD relies on extensive information exchange between agents, it is crucial for the underlying LLMs to support long-context windows and have strong instruction-following capabilities~\cite{liang-etal-2024-encouraging, du2023improving}. Therefore, we selected Qwen2.5-Coder~\cite{hui2024qwen2}, which ranks at the top of the Hugging Face open-source Code LLM Leaderboard~\cite{huggingfaceCodeModels}, with strong instruction-following capabilities, support for large context windows, and SOTA performance in code-related tasks and natural language inquiries.
For Code Summarization, we selected GPT-4o~\cite{gpt-4o} and CodeLlama-Instruct-7B~\cite{CodeLlama7bInstruct}, the best-performing proprietary and open-source LLMs used in the SOTA Code Summarization baseline~\cite{sun2024source}.

\subsection{Evaluation Metrics}\label{method_eval_metrics}

In this study, the outcome of the MAD debate is the final task output accepted by the judge agent, i.e., a predicted input/output, translated code, or code summary. We evaluate the correctness of the debate outcome against the ground-truth values provided by our selected datasets (Table~\ref{tab:task_summary}) using the task-specific metrics described below.

\subsubsection{\textbf{Code Input and Output Prediction}}\label{cr_metrics}

To evaluate the performance of Code Input Prediction and Code Output Prediction tasks, we report \textbf{\textit{pass@1}} rate, using the automated test cases provided by our selected datasets (Table~\ref{tab:task_summary}) since \textbf{\textit{pass@1}} is widely adopted in the prior work to access the correctness of input and output predictions~\cite{ding2024semcoder, chen2024reasoning, jain2025livecodebench, gu2024cruxeval}. All \textbf{\textit{pass@1}} scores are normalized to a 100 scale for readability.

\subsubsection{\textbf{Code Translation}}\label{ct_metrics}

To evaluate the performance of the Code Translation task, following the literature~\cite{jiao2023evaluation, zhu2024semi, pan2024lost, yang2024exploring, ahmad2023avatarparallelcorpusjavapython}, we report the \textbf{Computational Accuracy}, which indicates the percentage of generated programs that produce the correct output when executed on a given set of test cases. We used the automated test cases provided by our selected datasets (Table~\ref{tab:task_summary}). The scores are normalized to a 100 scale for readability.

\subsubsection{\textbf{Code Summarization}}\label{cs_metrics}
To evaluate the performance of the Code Summarization task, we employed a suite of evaluation metrics that capture both textual and semantic similarity between the generated summaries and human-written references. 
For textual similarity, we used widely adopted metrics in the literature that include \textbf{BLEU}, \textbf{METEOR}, and \textbf{ROUGE-L}~\cite{imani2024context, ahmed2024automatic, sun2024source, sun2024extractive}. 
To assess semantic similarity between generated and human-written summaries, we used widely adopted metrics in the literature~\cite{sun2024source, virk2025calibration, mastropaolo2024evaluating, haque2022semantic}, including \textbf{BERTScore}~\cite{zhang2020bertscoreevaluatingtextgeneration}, \textbf{SentenceBERT with Cosine Similarity (SBCS)}~\cite{reimers2019sentencebertsentenceembeddingsusing}, and \textbf{Universal Sentence Encoder with Cosine Similarity (USECS)}~\cite{cer2018universalsentenceencoder}. 
In addition, recent studies have emphasized the limitations of evaluation metrics based on n-gram overlap (i.e., BLEU) for assessing the quality of LLM-generated summaries~\cite{ahmed2024automatic, sun2024source, mastropaolo2024evaluating}. These studies have shown that human-written summaries are often incomplete or vary in quality, hence, relying solely on textual similarity measures may limit the assessment of the semantic quality captured in the summary~\cite{ahmed2024automatic}. To address this, we use \textbf{SIDE}~\cite{mastropaolo2024evaluating}, a recently proposed metric that directly measures semantic alignment between code and generated summaries, used in recent studies to evaluate the quality of LLM-generated summaries independent of human-written references~\cite{sun2024source, afrin2025resource}. All similarity scores range 0–1, with 1 indicating perfect alignment. All scores are normalized to 100 scale for readability.

\section{Results}
\label{section:result}

In this section, we organized the results of this study based on our research questions in Section \ref{section:intro}. \\ \\
\noindent\textbf{RQ1: How effective is Multi-Agent Debate when applied to coding tasks?}
The results for RQ1 are summarized in Tables~\ref{tab:cr_io_mad},~\ref{tab:ct_mad_all}, and~\ref{tab:cs_mad_result}, which report the performance of default MAD across the four coding tasks in SE using the evaluation metrics described in Section~\ref{method_eval_metrics}. To evaluate both the statistical and practical significance of the differences between default MAD and SOTA techniques, after checking for normality using Shapiro–Wilk test~\cite{shapiro_wilk}, we conducted Welch’s t-tests~\cite{ruxton2006unequal} and computed Cohen’s $d$ effect sizes~\cite{diener2010cohen}. Due to space limitations, the full statistical analysis, including all $p$-values and effect sizes, is available on our companion website~\cite{supply}. Asterisks (*) in Tables~\ref{tab:cr_io_mad},~\ref{tab:ct_mad_all}, and~\ref{tab:cs_mad_result} denote values with $p$-values below 0.05, indicating statistical significance.

\captionsetup{skip=1pt}
\hypertarget{tab:cr_io_mad}{}
\begin{table*}[htbp]
\footnotesize
\renewcommand{\arraystretch}{0.9}
\setlength{\tabcolsep}{2pt}
\centering
\caption{
\centering
Performance of MAD on Code Input and Output Prediction (pass@1)\\
\scriptsize \textit{Bold indicates the best performance. Asterisks (*) mark positive statistical significance (p-value < 0.05) compared to the SE Baseline.}}
\label{tab:cr_io_mad}
\resizebox{0.75\linewidth}{!}{
\begin{tabular}{lllccc}
\toprule
\multicolumn{1}{c}{\textbf{Model}} & \multicolumn{1}{c}{\textbf{Paradigm}} & \multicolumn{1}{c}{\textbf{Method}}
& \multicolumn{1}{c}{\textbf{Input Prediction}} 
& \multicolumn{2}{c}{\textbf{Output Prediction}} \\
\cmidrule(lr){4-4} \cmidrule(lr){5-6}
& & & \textbf{CRUXEval} 
& \textbf{CRUXEval} & \textbf{LiveCodeBench} \\
\midrule
\multirow{6}{*}{DeepSeek-Coder}
& \multirow{3}{*}{Single-Agent}
& Chain-of-Thought            & 31.8 & 29.2 & 26.1 \\
& & Self-Consistency         & 33.6 & 59.0 & 42.8 \\
& & SemCoder (SE Baseline)                      & 62.5 & 65.1 & 59.7 \\
\cmidrule(lr){2-6}
& \multirow{3}{*}{Multi-Agent}
& Default MAD                                   & 31.9 & 43.5 & 41.8 \\
& & Early Termination                           & 42.6 & 55.1 & \textbf{92.1$^{*}$} \\
& & Extended Reflection                         & \textbf{64.5$^{*}$} & \textbf{67.4$^{*}$} & 84.3$^{*}$ \\
\midrule
\multirow{6}{*}{GPT-4o}
& \multirow{3}{*}{Single-Agent}
& Chain-of-Thought           & 34.5 & 15.8 & 9.9 \\
& & Self-Consistency          & 34.7 & 15.1 & 9.5 \\
& & SemCoder (SE Baseline)                      & 69.3 & 73.6 & \textbf{65.3} \\
\cmidrule(lr){2-6}
& \multirow{3}{*}{Multi-Agent}
& Default MAD                                   & 56.4 & 27.8 & 50.1 \\
& & Early Termination                           & 46.8 & 57.2 & 47.4 \\
& & Extended Reflection                         & \textbf{73.5$^{*}$} & \textbf{93.2$^{*}$} & 48.4 \\
\bottomrule
\end{tabular}
}
\end{table*}

% TOSEM revision

% \captionsetup{skip=1pt}
% \begin{table*}[ht]
% \footnotesize
% \renewcommand{\arraystretch}{0.5}
% \setlength{\tabcolsep}{2pt}
% \centering
% \caption{
% \centering
% Performance of MAD on Code Input and Output Prediction (pass@1)\\
% \scriptsize \textit{Bold indicates the best performance. Asterisks (*) mark statistical significance (p-value < 0.05)}}
% \label{tab:cr_io_mad}
% \resizebox{0.75\linewidth}{!}{
% \begin{tabular}{llccc}
% \toprule
% \textbf{Model} & \textbf{Method} 
% & \multicolumn{1}{c}{\textbf{Input Prediction}} 
% & \multicolumn{2}{c}{\textbf{Output Prediction}} \\
% \cmidrule(lr){3-3} \cmidrule(lr){4-5}
% & & \textbf{CRUXEval} 
%   & \textbf{CRUXEval} & \textbf{LiveCodeBench} \\
% \midrule
% \multirow{4}{*}{DeepSeek-Coder}
% & SemCoder (Baseline)     & 62.5 & 65.1 & 59.7 \\
% & Default MAD             & 31.9$^{*}$ & 43.5$^{*}$ & 41.8$^{*}$ \\
% & Early Termination       & 42.6$^{*}$ & 55.1$^{*}$ & \textbf{92.1$^{*}$}   \\
% & Extended Reflection     & \textbf{64.5} & \textbf{67.4} & 84.3$^{*}$ \\
% \midrule
% \multirow{4}{*}{GPT-4o}
% & SemCoder (Baseline)     & 69.3 & 73.6 & \textbf{65.3} \\
% & Default MAD             & 56.4$^{*}$ & 27.8$^{*}$ & 50.1$^{*}$ \\
% & Early Termination       & 46.8$^{*}$ & 57.2$^{*}$ & 47.4$^{*}$  \\
% & Extended Reflection     & \textbf{73.5} & \textbf{93.2$^{*}$} & 48.4$^{*}$  \\
% \bottomrule
% \end{tabular}
% }
% \end{table*}
% \input{tables/ct_mad_all}
% \input{tables/cs_mad_result}

For the Code Input Prediction task, the results in Table~\ref{tab:cr_io_mad} show that Default MAD performs comparable to the two Single-Agent SOTA baselines and performs worse than the SE SOTA baseline, SemCoder~\cite{ding2024semcoder}, across both models. This suggests that Default MAD has difficulty generating valid input values that satisfy a program’s input constraints when conditioned on its output.
Similarly, for Code Output Prediction, the results in Table~\ref{tab:cr_io_mad} show that Default MAD outperforms or achieves comparable performance to the Single-Agent baselines. However, the SE SOTA baseline outperforms Default MAD across both models and datasets, with small effect sizes (DeepSeek on CruxEval: Cohen's d= 0.42, a small effect size). These results suggest that the Default MAD falls short in predicting the input and output of the program that is functionally correct and satisfies the program constraints.

% This indicates that Default MAD struggles to produce functionally correct translations consistently. 
% Nevertheless, when considering Execution Accuracy—the percentage of translated code that runs without compilation or runtime errors—Default MAD outperforms PLTranslation (p-value < 0.05, Cohen's $d$ = 0.41, medium effect size). 
% These results suggest that while Default MAD may produce syntactically correct and executable code more frequently, it is less effective at generating semantically and functionally accurate translations. 

\captionsetup{skip=1pt}
\begin{table*}[htbp]
\footnotesize
\renewcommand{\arraystretch}{0.9}
\setlength{\tabcolsep}{2pt}
\centering
\caption{
\centering
Performance of MAD on Code Translation.\\
\scriptsize \textit{Bold indicates the best performance. Asterisks (*) mark statistical significance (p-value < 0.05)}}
\label{tab:ct_mad_all}
\resizebox{0.75\linewidth}{!}{
\begin{tabular}{lll cc cc cc}
\toprule
\multicolumn{1}{c}{\textbf{Model}} & \multicolumn{1}{c}{\textbf{Paradigm}} & \multicolumn{1}{c}{\textbf{Method}}
& \multicolumn{2}{c}{\textbf{CodeNet}} 
& \multicolumn{2}{c}{\textbf{AVATAR}} 
& \multicolumn{2}{c}{\textbf{EvalPlus}} \\
\cmidrule(lr){4-5} \cmidrule(lr){6-7} \cmidrule(lr){8-9}
& & & \textbf{Python} & \textbf{Java}
  & \textbf{Python} & \textbf{Java}
  & \textbf{Python} & \textbf{Java} \\
\midrule
\multirow{6}{*}{Qwen}
& \multirow{3}{*}{Single-Agent}
& Chain-of-Thought & 42.79 & 0 & 51.29 & 0 & 61.43 & 0 \\
& & Self-Consistency & 55.31 & 0 & 56.94 & 0 & 60.75 & 0 \\
& & PLTranslation (SE Baseline)      & 69.50 & 59.20 & 46.00 & 66.67 & 23.17 & 23.41 \\
\cmidrule(lr){2-9}
& \multirow{3}{*}{Multi-Agent}
& Default MAD         & 62.50 & 56.22 & 51.60 & 49.00 & 26.22 & 50.61$^{*}$ \\
& & Early Termination   & 59.40 & \textbf{63.00} & \textbf{67.20}$^{*}$ & 59.36 & 62.20 & 74.39$^{*}$ \\
& & Extended Reflection & \textbf{75.43} & 62.02 & 64.40$^{*}$ & \textbf{69.47} & \textbf{70.12}$^{*}$ & \textbf{96.34}$^{*}$ \\
\midrule
\multirow{6}{*}{GPT-4o}
& \multirow{3}{*}{Single-Agent}
& Chain-of-Thought & 46.17 & 0 & 44.70 & 0 & 69.50 & 0 \\
& & Self-Consistency & 41.00 & 0 & 46.18 & 0 & 65.22 & 0 \\
& & PLTranslation (SE Baseline)      & 51.50 & 26.00 & 39.76 & 21.24 & 21.34 & 33.54 \\
\cmidrule(lr){2-9}
& \multirow{3}{*}{Multi-Agent}
& Default MAD         & 43.41 & 23.09 & 38.40 & 23.65 & 25.00 & 68.90$^{*}$ \\
& & Early Termination   & 40.14 & 33.72 & 41.80 & 34.90$^{*}$ & 40.24 & 74.39$^{*}$ \\
& & Extended Reflection & \textbf{52.09} & \textbf{50.00}$^{*}$ & \textbf{50.80}$^{*}$ & \textbf{65.86}$^{*}$ & \textbf{46.95} & \textbf{80.48}$^{*}$ \\
\bottomrule
\end{tabular}
}
\end{table*}

For Code Translation, the results in Table~\ref{tab:ct_mad_all} highlight a critical limitation of Single-Agent baselines. In specific, Java-to-Python translation tasks using the baseline prompting methods (CoT and SC) failed to generate any executable code, resulting in a 0\% pass rate across all test cases. Compared to the SE SOTA baseline, PLTranslation~\cite{pan2024lost}, Default MAD achieves comparable or slightly lower performance on the CodeNet and AVATAR datasets. However, the differences observed are either not statistically significant or exhibit a small effect size (e.g., Qwen on AVATAR for Java, Cohen’s $d = 0.35$). In contrast, Default MAD outperforms the SOTA baseline on the EvalPlus dataset with statistical significance (Qwen on EvalPlus for Java, Cohen’s $d = 0.57$, medium effect size). These results suggest that while Default MAD achieves comparable performance on some datasets, it does not consistently outperform the SOTA SE baseline in accurately translating code functionality.

\captionsetup{skip=1pt}
\begin{table*}[htbp]
\footnotesize
\renewcommand{\arraystretch}{0.9}
\setlength{\tabcolsep}{2pt}
\newcommand{\highlight}[1]{\fbox{#1}}
\centering
\caption{
\centering
Performance of MAD on Code Summarization. \\
\scriptsize \textit{Bold indicates the best performance. Asterisks (*) mark statistical significance (p-value < 0.05)}}
\label{tab:cs_mad_result}
\resizebox{0.95\linewidth}{!}{
\begin{tabular}{ccllcccccccc}
    \toprule
    \textbf{Language} & \textbf{Model} & \multicolumn{1}{c}{\textbf{Paradigm}} & \multicolumn{1}{c}{\textbf{Method}} & \multicolumn{3}{c}{\textbf{Textual Similarity}} & \multicolumn{3}{c}{\textbf{Semantic Similarity}} & \textbf{Summary-Code} \\
    \cmidrule(lr){5-7} \cmidrule(lr){8-11} \cmidrule(lr){12-12}
    & & & & BLEU & METEOR & ROUGE-L & BERTScore & SBCS & USECS & SIDE \\
    \midrule
    \multirow{8}{*}{Python}
    & \multirow{4}{*}{CodeLLaMA}
    & \multirow{3}{*}{Single-Agent}
    & Chain-of-Thought & 6.77 & 10.45 & 11.88 & 42.67 & 47.39 & 30.96 & 49.35 \\
    & & & Self-Consistency & 6.71 & 10.13 & 11.57 & 52.64 & 47.23 & 30.97 & 48.36 \\
    & & & ASCS (SE Baseline) & 9.19 & 17.24 & 11.32 & 48.30 & 69.45 & 37.82 & 36.45 \\
    \cmidrule(lr){3-12}
    & & \multirow{3}{*}{Multi-Agent}
    & Default MAD & 13.39$^{*}$ & 20.04$^{*}$ & 16.45$^{*}$ & 52.46$^{*}$ & 75.90$^{*}$ & 42.98$^{*}$ & 90.47$^{*}$ \\
    & & & Early Termination & 13.74$^{*}$ & 20.67$^{*}$ & 17.02$^{*}$ & 52.63$^{*}$ & 76.15$^{*}$ & 43.58$^{*}$ & \textbf{92.12}$^{*}$ \\
    & & & Extended Reflection & \textbf{16.32}$^{*}$ & \textbf{21.79}$^{*}$ & \textbf{22.64}$^{*}$ & \textbf{54.80}$^{*}$ & \textbf{80.25}$^{*}$ & \textbf{44.71}$^{*}$ & 92.10$^{*}$ \\
    \cmidrule(lr){2-12}
    & \multirow{4}{*}{GPT-4o}
    & \multirow{3}{*}{Single-Agent}
    & Chain-of-Thought & 7.02 & 19.86 & 18.06 & 42.46 & 48.38 & 31.11 & 49.92 \\
    & & & Self-Consistency & 6.29 & 19.68 & 17.59 & 42.48 & 46.66 & 39.95 & 49.25 \\
    & & & ASCS (SE Baseline) & 15.80 & 20.01 & 20.78 & 47.02 & 51.43 & 44.22 & 45.63 \\
    \cmidrule(lr){3-12}
    & & \multirow{3}{*}{Multi-Agent}
    & Default MAD & 18.48$^{*}$ & 23.95$^{*}$ & 19.59$^{*}$ & 51.89$^{*}$ & 75.64$^{*}$ & 45.40 & 89.83$^{*}$ \\
    & & & Early Termination & 21.67$^{*}$ & 22.56$^{*}$ & 16.72 & 52.48$^{*}$ & 74.78$^{*}$ & 45.00 & \textbf{92.43}$^{*}$ \\
    & & & Extended Reflection & \textbf{23.87}$^{*}$ & \textbf{25.03}$^{*}$ & \textbf{24.89}$^{*}$ & \textbf{56.79}$^{*}$ & \textbf{81.02}$^{*}$ & \textbf{48.72}$^{*}$ & 92.06$^{*}$ \\
    \midrule
    \multirow{8}{*}{Java}
    & \multirow{4}{*}{CodeLLaMA}
    & \multirow{3}{*}{Single-Agent}
    & Chain-of-Thought & 6.82 & 12.53 & 14.61 & 33.70 & 32.49 & 35.67 & 38.37 \\
    & & & Self-Consistency & 7.07 & 12.45 & 14.56 & 33.52 & 31.74 & 35.69 & 38.77 \\
    & & & ASCS (SE Baseline) & 13.00 & 17.54 & 15.05 & 50.43 & 74.93 & 38.51 & 34.70 \\
    \cmidrule(lr){3-12}
    & & \multirow{3}{*}{Multi-Agent}
    & Default MAD & 14.32 & 16.92 & 16.22 & 50.69 & 76.20$^{*}$ & 38.89  & 89.23$^{*}$ \\
    & & & Early Termination & \textbf{14.34} & \textbf{18.36} & 16.61 & 50.87 & 75.90 & \textbf{39.94}$^{*}$ & \textbf{90.66}$^{*}$ \\
    & & & Extended Reflection & 13.87 & 17.85 & \textbf{19.27}$^{*}$ & \textbf{51.70}$^{*}$ & \textbf{79.08}$^{*}$ & 39.46$^{*}$ & 90.64$^{*}$ \\
    \cmidrule(lr){2-12}
    & \multirow{4}{*}{GPT-4o}
    & \multirow{3}{*}{Single-Agent}
    & Chain-of-Thought & 6.81 & 11.95 & 15.61 & 33.37 & 32.41 & 34.89 & 39.11 \\
    & & & Self-Consistency & 6.73 & 11.83 & 15.55 & 33.41 & 32.11 & 33.92 & 38.63 \\
    & & & ASCS (SE Baseline) & 14.83 & 17.04 & 19.94 & 47.46 & 52.27 & 38.61 & 41.23 \\
    \cmidrule(lr){3-12}
    & & \multirow{3}{*}{Multi-Agent}
    & Default MAD & 16.24$^{*}$ & 19.18$^{*}$ & 19.57$^{*}$ & 45.33 & 70.84$^{*}$ & 38.08 & 89.50$^{*}$ \\
    & & & Early Termination & 20.33$^{*}$ & 19.57$^{*}$ & 16.50 & 47.59 & 71.37$^{*}$ & \textbf{43.89}$^{*}$ & 90.24$^{*}$ \\
    & & & Extended Reflection & \textbf{23.17}$^{*}$ & \textbf{21.39}$^{*}$ & \textbf{22.08}$^{*}$ & \textbf{53.51}$^{*}$ & \textbf{79.83}$^{*}$ & 42.73$^{*}$ & \textbf{90.39}$^{*}$ \\
    \bottomrule
\end{tabular}
}
\end{table*}

For Code Summarization, the evaluation metrics (Section~\ref{cs_metrics}) assess task performance from multiple perspectives. On metrics that measure textual and semantic similarity between generated and human-written summaries, the Default MAD generally matches or outperforms the performance of the Single-Agent Prompting baselines and SE SOTA baseline, ASCS with a medium effect size (CodeLLaMA for Python using ROUGE-L Cohen's $d$ = 0.51, medium effect size). However, since human-written summaries can often be of low quality \cite{linares2015developers, wen2019large}, such similarity-based metrics may not fully reflect the quality of generated summaries. To address this, we used a diverse set of metrics (Section~\ref{method_eval_metrics}) that measures textual, semantic, and summary-code similarity for a more comprehensive evaluation. For instance, SIDE, which directly measures semantic alignment between the code and generated summaries, shows that Default MAD statistically significantly outperforms ASCS with a large effect size (CodeLLaMA for Python using SIDE Cohen's $d$ = 5.40, large effect size). Overall, these results indicate that Default MAD, though designed for general NLP tasks, can be effectively adapted to Code Summarization to produce summaries that better capture the functionality of source code.

\begin{tcolorbox}[
  colback=white,
  boxsep=1pt,       % tighter internal padding
  left=2pt,         % left padding
  right=2pt,        % right padding
  top=2pt,          % top padding
  bottom=2pt,       % bottom padding
]
\textbf{Observation 1:} The default MAD, designed for general NLP tasks, can be effectively applied to Code Summarization. However, it yields comparable performance in Code Translation and shows significant underperformance in Code Input and Output Prediction.
\end{tcolorbox}

% \vspace{5pt}
\noindent\textbf{RQ2: What are the debate patterns in MAD that are responsible for the underperforming cases of coding tasks?}\\
In RQ2, we aim to improve the performance of the Default MAD on the selected coding tasks by analyzing its underperforming cases. Specifically, we manually analyzed the debate logs of such cases to identify the potential patterns that may be responsible for the underperformance  (Section~\ref{method_debate_cateogries}). The debate pattern categories are listed as follows:

\begin{itemize}[label=•, leftmargin=10pt]\hypertarget{debate_categories}{}
\setlength{\itemsep}{2pt}
\setlength{\parskip}{0pt}
\item \textbf{Forceful Agreement}: The \textit{Position}s of the majority of agents are \textit{Disagree} at the beginning of the debate, but change to \textit{Agree} at the end. This category follows a pattern where an agent with an initially acceptable answer is negatively impacted by other agents, as low-quality responses \textit{``force''} the debate toward an incorrect consensus.

\item \textbf{Ending Divergence}: 
The \textit{Position}s of the majority of agents are initially \textit{Agree} at the start of the debate but shift to \textit{Disagree} by the end. This category captures cases where agents begin by agreeing on an acceptable response but gradually diverge during the debate, ultimately arriving at responses deemed unacceptable by the judge.

Un\item \textbf{Prolonged Disagreement}: The \textit{Position}s of the majority of the agents are \textit{Disagree} at the beginning of the debate, and remain as \textit{Disagree} at the end. In this case, the agents resist being influenced by other agents' responses. Hence, agents do not iteratively agree and refine each other's responses, causing the quality of the response to be stagnant, given the initial response is of low quality or not acceptable to the judge. 
\end{itemize}

\captionsetup{skip=1pt}
\begin{table}[h]
\footnotesize
\setlength{\tabcolsep}{2pt}
\centering
\caption{
\centering
Underperforming Debate Categorization Result\\
\scriptsize \textit{Bold indicate most prevalent category. (260 samples, 87.3\% IRR)}}
\resizebox{0.95\linewidth}{!}{
\begin{tabular}{l@{\hspace{0.8cm}}c@{\hspace{0.6cm}}c@{\hspace{0.4cm}}c@{\hspace{0.4cm}}c@{\hspace{0.4cm}}c@{\hspace{0.4cm}}c}
\toprule
\multirow{2}{*}{\textbf{Debate Category}} & \multirow{2}{*}{\textbf{Human}} & \multicolumn{5}{c}{\textbf{LLM}} \\
\cmidrule(lr){3-7}
& & \textbf{CIP} & \textbf{COP} & \textbf{CT} & \textbf{CS} & \textbf{All Tasks} \\
\midrule
Forceful Agreement         & 20 (7.69\%)     & 14 (14\%)       & 36 (20.45\%)     & 31 (7.19\%)     & 21 (26.58\%)     & 102 (12.98\%) \\
Ending Divergence          &  \textbf{199 (76.54\%)}   & 20 (20\%)       & 52 (29.55\%)     & \textbf{343 (79.58\%)}   & \textbf{50 (63.29\%)}     & \textbf{465 (59.14)} \\
Prolonged Disagreement     & 41 (15.77\%)    & \textbf{66 (66\%)}      & \textbf{88 (50\%)}       & 57 (13.23\%)    & 8 (10.13\%)      & 219 (27.89\%) \\
\midrule
\# of Samples      & 260            & 100 (12.50\%)            & 176 (13.76\%)            & 431 (35.09\%)            & 79 (5.94\%)             & 786 (16.95\%) \\
\bottomrule
\end{tabular}
}
\label{tab:rq2_categorization}
\end{table}

Table~\ref{tab:rq2_categorization} shows the frequency of each debate pattern, categorized by human annotators and LLM-generated results using GPT-4o~\cite{gpt-4o}. The \textbf{Human} column representing human annotation and the \textbf{LLM} column representing LLM annotation across all tasks exhibit a similar distribution of debate categories. The last row \textbf{Underperforming Debate}, reports the count and percentage of underperforming debate cases out of the total samples used in our experiment for each task (i.e. for the CIP task, 12.5\% corresponds to 100 underperforming cases out of 800 data samples).

Table~\ref{tab:rq2_complexity} presents the code complexity measures for each underperforming debate category across all tasks, comparing them against the \textbf{Successful} debate (debate cases with a valid task outcome determined by the judge agent). The columns report the average number of statements (\textbf{\# Stmts}), operands (\textbf{\# Oprnd}), Cyclomatic Complexity (\textbf{CC}), average nested block depth (\textbf{NestDpth}), and average number of parameters (\textbf{\# Params}) per category and coding task included in our study (details on code complexity measures in Section~\ref{complexity_metrics}). Notably, the \textbf{Successful} category exhibits a similar complexity measures across all three underperforming debate categories namely, \textbf{Forceful Agreement}, \textbf{Ending Divergence}, and \textbf{Prolonged Disagreement} implying that code complexity alone is not a reliable indicator for distinguishing underperforming debates from successful ones. Nevertheless, we observe that the CT (Java to Python translation) task consistently exhibits higher complexity values compared to the other coding tasks (CIP, COP, and CS) across all debate categories. For example, \textbf{\# Stmts}, \textbf{\# Oprnd}, and \textbf{CC} are consistently higher for CT tasks, suggesting that code samples in translation tasks tend to be structurally more complex, potentially posing a greater challenge for agents to reason about and resolve effectively.

From the Table~\ref{tab:rq2_categorization}, we observe notable differences in underperformance patterns between the tasks. For instance, in the Code Input Prediction and Code Output Prediction tasks, the most prevalent category was \textit{Prolonged Disagreement} (debates where initial disagreement persisted until the end). In contrast, for Code Summarization and Code Translation, the most dominant category was \textit{Ending Divergence} (debates that began with agreement but ended in disagreement). This suggests that divergence emerging toward the end of a debate may contribute to failed debate outcomes, as it makes it more difficult for the judge to identify a clear winning agent.

This pattern may also reflect differences in task output. For instance, Code Summarization and Code Translation allow for multiple valid responses~\cite{improvesource2018, pan2024lost}, where multiple summaries or alternative code translations can exist for a given code. This allows for variability in responses, which can hinder alignment between agents and reduce the likelihood of reaching a consensus, leading to underperformance. In contrast, Code Input Prediction and Code Output Prediction require precise outputs that are constrained by program specifications (i.e., input types and function signatures)~\cite{gu2024cruxeval, jain2025livecodebench}, limiting the variability of acceptable answers. Hence, the disagreements between agents are more persistent when responses diverge, resulting in underperformance due to a failure to reach consensus.

% Table: Mean code complexity metrics per debate category, all tasks.
% Underperforming rows: CIP/COP from CRUXEval (Python, deduped by sample);
%   CT from Java source; CS from Python source.
% Normal (non-underperforming) row uses samples not assigned any debate
%   underperformance category in the same dataset.

\begin{table*}[t]
\centering
\small
\caption{\centering Code Complexity of Underperforming Debate Categories.\vspace{2pt}
\scriptsize (Successful = non-underperforming debate)\\
Stmts = Statements; Oprnd = Operands; CC = Cyclomatic Complexity;
NestDpth = Avg.\ Nested Block Depth; Params = Avg.\ Parameters.}
\label{tab:rq2_complexity}
\begin{tabular}{@{} l l rrrrr @{}}
\toprule
\textbf{Category} & \textbf{Task} & \textbf{\# Stmts} & \textbf{\# Oprnd} & \textbf{CC} & \textbf{NestDpth} & \textbf{\# Params} \\
\midrule
  \multirow{4}{*}{Forceful Agreement}
   & CIP (Python)  & 10.8 & 24.4 & 3.4 & 0.3 & 0.9 \\
   & COP (Python)  & 10.7 & 24.7 & 3.3 & 0.3 & 0.9 \\
   & CS (Python)   & 11.4 & 29.4 & 4.1 & 2.5 & 2.6 \\
   & CT (Java)     & 20.0 & 44.8 & 4.8 & 2.1 & 1.0 \\
\cmidrule(lr){1-7}
  \multirow{4}{*}{Ending Divergence}
   & CIP (Python)  & 10.7 & 24.0 & 3.3 & 0.3 & 0.9 \\
   & COP (Python)  & 11.2 & 25.3 & 3.5 & 0.4 & 0.8 \\
   & CS (Python)   & 11.6 & 34.8 & 3.1 & 2.3 & 2.6 \\
   & CT (Java)     & 24.2 & 53.1 & 6.6 & 2.4 & 1.0 \\
\cmidrule(lr){1-7}
  \multirow{4}{*}{Prolonged Disagreement}
   & CIP (Python)  & 11.7 & 28.3 & 3.8 & 0.4 & 0.9 \\
   & COP (Python)  & 10.9 & 26.1 & 3.1 & 0.3 & 0.9 \\
   & CS (Python)   & 10.8 & 26.4 & 2.8 & 2.1 & 2.1 \\
   & CT (Java)     & 23.3 & 54.4 & 6.0 & 2.4 & 1.0 \\
\cmidrule(lr){1-7}
  \multirow{4}{*}{Successful}
   & CIP (Python)  & 10.7 & 24.1 & 3.2 & 0.3 & 0.8 \\
   & COP (Python)  & 10.6 & 23.7 & 3.3 & 0.3 & 0.8 \\
   & CS (Python)   & 12.3 & 33.6 & 4.3 & 2.4 & 2.5 \\
   & CT (Java)     & 22.2 & 49.6 & 6.0 & 2.3 & 1.1 \\
\bottomrule
\end{tabular}
\end{table*}

\begin{tcolorbox}[
  colback=white,
  boxsep=1pt,       % tighter internal padding
  left=2pt,         % left padding
  right=2pt,        % right padding
  top=2pt,          % top padding
  bottom=2pt,       % bottom padding
]
\textbf{Observation 2:} Debate patterns that cause underperformance vary by task but are rooted in divergence in responses that lead to unresolved disagreement.
\end{tcolorbox}

% para 5
Following the underperforming debate categorization, we further analyze the root causes behind each category through manual analysis of the debate logs. Specifically, we examined debate logs in each category to identify the underlying causes that led to debate failures, including forceful agreement, ending divergence, and prolonged disagreement among agents engaged in MAD. Table~\ref{tab:root_cause_def} presents the root cause definitions for each debate failure category. We identified both category-specific root causes and shared root causes prevalent across all debate categories. The details on how these root causes were identified are outlined in Section~\ref{root_cause_methods}. 

\begin{table}[htbp]
\footnotesize
\centering
\captionsetup{skip=1pt}
\caption{Root Cause Definitions for Underperforming Debate Categories.}
\label{tab:root_cause_def}
\begin{tabular}{llp{7cm}}
\toprule
\textbf{Category} & \textbf{Root Cause} & \textbf{Description} \\
\midrule
\multirow{2}{*}{\shortstack[l]{Forceful\\Agreement}} & 
Shared Hallucination (SH) & Agents share the same incorrect assumption before the debate begins, forming a false consensus that leaves the error undetected. \\[12pt] & 
Peer Pressure (PP) & Agents converge on a wrong answer due to majority pressure rather than evidence or reasoning. \\[10pt]
\midrule
\multirow{2}{*}{\shortstack[l]{Ending\\Divergence}} &
Answer Drift (AD) & Agents start with the same answer but identify different faults across iterations, leading to different final answers. \\[12pt] &
Disagreement in Details (DD) & Agents agree on the high-level answer but diverge when elaborating on specific details. \\[10pt]
\midrule
\multirow{2}{*}{\shortstack[l]{Prolonged\\Disagreement}} & 
Fixed Stance (FS) & Agents persistently reassert their initial answers, reflecting a adherence to initial positions that prevents convergence. \\[12pt] &
Symmetric Hallucination (SyH) & Agents hold different but incorrect assumption, making convergence on the correct answer challenging. \\[10pt]
\midrule
\multirow{2}{*}{\shortstack[l]{Shared Across\\All Categories}} &
Judge Anchoring (JA) & The judge's selection in earlier rounds of iteration anchors agents to a potentially incorrect answer. \\[12pt] &
Context Decay (CD) & As debate iterations progress, agents increasingly rely on each other's responses rather than the original source, causing key contextual information to be gradually lost or distorted. \\[10pt]
\bottomrule
\end{tabular}
\end{table}
\begin{table*}[htbp]
\footnotesize
\centering
\captionsetup{skip=1pt}
\caption{\centering Root Causes of Underperforming Debates (786 samples).\\
\scriptsize Bold indicates the most prevalent root cause within each debate category. $^*$indicates the most prevalent root cause per task across all categories.}
\label{tab:root_cause}
\setlength{\tabcolsep}{9pt}
\begin{tabular}{llrrrrr}
\toprule
\textbf{Debate Category} & \textbf{Root Cause} & \textbf{CIP} & \textbf{COP} & \textbf{CT} & \textbf{CS} & \textbf{All Tasks} \\
\midrule
%% ---- Forceful Agreement (D-C) ----------------------------------------
\multirow{5}{*}{\shortstack[l]{Forceful\\Agreement}}
  & \textbf{Shared Hallucination (SH)}  &  7 & 18 & 15 & 10$^*$ & \textbf{50 (49.0\%)} \\
  & Peer Pressure (PP)                  &  4 &  8 &  8 &  5 &          25 (24.5\%) \\
  & Judge Anchoring (JA)               &  2 &  6 &  8 &  4 &          20 (19.6\%) \\
  & Context Decay (CD)                 &  1 &  4 &  0 &  2 &           7 (6.9\%)  \\
\cmidrule(lr){2-7}
  & \textit{Total per Category}
    & \textit{14} & \textit{36} & \textit{31} & \textit{21} & \textit{102} \\
\midrule
%% ---- Ending Divergence (C-D) ------------------------------------------
\multirow{5}{*}{\shortstack[l]{Ending\\Divergence}}
  & \textbf{Answer Drift (AD)}          &  6 & 17 & 120$^*$ & 16 & \textbf{159 (34.2\%)} \\
  & Disagreement in Details (DD)        &  5 & 13 &  98 & 12 &          128 (27.5\%) \\
  & Judge Anchoring (JA)               &  5 & 12 &  78 & 12 &          107 (23.0\%) \\
  & Context Decay (CD)                 &  4 & 10 &  47 & 10 &           71 (15.3\%) \\
\cmidrule(lr){2-7}
  & \textit{Total per Category}
    & \textit{20} & \textit{52} & \textit{343} & \textit{50} & \textit{465} \\
\midrule
%% ---- Prolonged Disagreement (D-D) -------------------------------------
\multirow{5}{*}{\shortstack[l]{Prolonged\\Disagreement}}
  & \textbf{Fixed Stance (FS)}          & 24$^*$ & 30$^*$ & 20 &  3 & \textbf{77 (35.2\%)} \\
  & Symmetric Hallucination (SyH)       & 20 & 25 & 18 &  2 &          65 (29.7\%) \\
  & Judge Anchoring (JA)               & 13 & 18 & 12 &  2 &          45 (20.5\%) \\
  & Context Decay (CD)                 &  9 & 15 &  7 &  1 &          32 (14.6\%) \\
\cmidrule(lr){2-7}
  & \textit{Total per Category}
    & \textit{66} & \textit{88} & \textit{57} & \textit{8} & \textit{219} \\
\midrule
%% ---- Total ------------------------------------------------------------
\multicolumn{2}{l}{\textbf{Total}}
  & \textbf{100} & \textbf{176} & \textbf{431} & \textbf{79} & \textbf{786} \\
\bottomrule
\end{tabular}
\end{table*}

Table~\ref{tab:root_cause} presents the frequency of each root cause across the 786 underperforming debate logs. To detail, the identified root causes for the \textit{Forceful Agreement} debate category include \textit{Shared Hallucination} and \textit{Peer Pressure}. In \textit{Shared Hallucination}, agents share incorrect assumptions about the task, such as the execution environment, contextual information about the code, or implicit import statements. Despite the presence of an initially correct answer, these incorrect assumptions cause agents to converge on an incorrect answer as debate iterations progress. In \textit{Peer Pressure}, agents change their correct answers upon observing the majority opinion from other agents, rather than seeking further validation or defending their reasoning, leading the debate to converge on an incorrect conclusion. As shown in Table~\ref{tab:root_cause}, \textit{Shared Hallucination} is the dominant root cause (49.0\%) in this category. This indicates that incorrect convergence is largely driven by unchallenged shared assumptions, highlighting the need for intervention mechanisms that explicitly validate agents’ intermediate answers to support effective debate.

For the \textit{Ending Divergence} category, the root causes include \textit{Answer Drift}, where agents begin aligned but progressively identify different faults across iterations, and \textit{Disagreement in Details}, where agents agree on the high-level answer but diverge on specific details. Table~\ref{tab:root_cause} shows that \textit{Answer Drift} is the most common root cause (34.2\%) in this category. This result suggests that the iterative nature of debate can be counterproductive, as agents that begin aligned may gradually introduce inconsistencies through successive refinements. For tasks with multiple valid or acceptable answers, defining an appropriate \textbf{stopping mechanism} is important to prevent the debate from leading to suboptimal outcomes.

For the \textit{Prolonged Disagreement} category, the root causes are \textit{Fixed Stance} and \textit{Symmetric Hallucination}. In \textit{Fixed Stance}, agents persistently reassert their initial answers despite counter-arguments or alternative solutions, without providing additional rationale, making convergence difficult. In \textit{Symmetric Hallucination}, agents hold different but equally incorrect assumptions about the task, causing the debate to focus on resolving these discrepancies rather than progressing toward a correct solution. Table~\ref{tab:root_cause} shows that \textit{Fixed Stance} is the most prevalent root cause (35.2\%) in this category, indicating that the debate mechanism fails to facilitate meaningful exchange, with agents repeating their positions. This suggests the need to implement a guided mechanism to \textbf{reset the ground} in MAD, enabling the debate to recover and restore meaningful interaction among agents.

Finally, \textit{Judge Anchoring} and \textit{Context Decay} are prevalent root causes across all debate categories. In \textit{Judge Anchoring}, the answer selected by the judge in earlier rounds influences subsequent debate iterations, steering agents toward a judge-preferred answer. In \textit{Context Decay}, agents increasingly focus on critiquing each other’s responses, causing key contextual information to be gradually lost or distorted. We observe that these root causes occur in approximately 20\% of cases for \textit{Judge Anchoring} and around 10\% for \textit{Context Decay} across all debate categories (Table~\ref{tab:root_cause}). This suggests that moderating what agents observe is necessary to avoid overly influencing their decisions, and that minimizing unnecessary interactions between agents may improve outcomes. Additionally, longer debates do not necessarily yield better results, as maintaining concise and focused debates while reinforcing the original context may help mitigate context decay and improve the quality of agent interactions.

Notably, for CIP and COP, \textit{Fixed Stance} is the most prevalent root cause, whereas for CT, \textit{Answer Drift} is the dominant root cause. This aligns with earlier observations from Table~\ref{tab:rq2_categorization}, where CIP and COP tasks require precise outputs constrained by program specifications, which may lead agents to persistently reassert their initial answers even within structured argumentation. In contrast, CT is more susceptible to \textit{Answer Drift}, as multiple valid answers exist for solving the same problem. For example, a looping algorithm can be implemented in various ways when translating across programming languages, causing agents to focus on different aspects of the solution across iterations. For CS, \textit{Shared Hallucination} is the most prevalent root cause, suggesting that tasks requiring rich contextual understanding such as function interactions, execution environments, and necessary imports, are particularly sensitive to assumptions or prior knowledge embedded in the backbone LLM used for building agents in MAD.

\textbf{RQ3: How can we improve the performance of MAD on coding tasks?}\label{sec:res_rq3}

The bottom two rows in Tables~\ref{tab:cr_io_mad}, \ref{tab:ct_mad_all}, and \ref{tab:cs_mad_result} present the evaluation results of applying our two MAD enhancement strategies (Section~\ref{sec:method_rq3}). We also conducted Welch’s t-test~\cite{ruxton2006unequal} and Cohen's $d$ \cite{diener2010cohen}. Due to space limitations, the full statistical analysis is available on our companion website~\cite{supply}.

%Table~\ref{tab:rq3_winner} shows the count of best-performing techniques per task based on model–dataset combinations. Each value represents the number of model–dataset combinations in which the corresponding technique outperformed all others for that specific task. For example, the value 2 in the CIP column for \textit{Extended Reflection} indicates that this technique achieved the highest performance on the CRUXEval dataset using two models (DeepSeek and GPT-4o) for the Code Input Prediction task. 

% Asterisks (*) in Tables~\ref{tab:cr_io_mad},~\ref{tab:ct_mad_all}, and~\ref{tab:cs_mad_result} denote values with $p$-values below 0.05, indicating statistical significance.

% I/O: two MAD variants vs SOTA 

% 1. default mad did not outperform sota
% 2. early termination still fall short but improve upon default mad
% 3. extended reflection outperform sota and improve upon default mad
For Code Input Prediction, \textit{Early Termination} showed limited performance gain compared to the SOTA baseline(DeepSeek Cohen's $d$ = 0.42). In contrast, \textit{Extended Reflection} outperformed the SOTA baseline~\cite{ding2024semcoder}, but the performance difference was not statistically significant (p-value > 0.05). Similarly, for Code Output Prediction, \textit{Early Termination} achieves lower performance compared to the SOTA baseline (GPT-4o on CRUXEval Cohen's $d$ = 0.34). However, the \textit{Extended Reflection} outperforms the SOTA baseline with statistical significance and a medium effect size (GPT-4o on CRUXEval, Cohen's $d$ = 0.54). 

% Table~\ref{tab:rq3_winner} shows that for Code Input Prediction and Output Prediction \textit{Extended Reflection} achieves the best performance overall.

% Prolonged Disagreement is when varied answers are present in the debate, and the agents in the debate fail to reach convergence through the process. 

% I/O: two MAD variants vs Default MAD
%OLD:This suggests that \textit{Early Termination} was not effective in mitigating underperforming cases associated with the \textit{Prolonged Disagreement} pattern (Table~\ref{tab:rq2_categorization}), which was most prevalent in the Code Input Prediction and Output Prediction tasks. \textit{Prolonged Disagreement} refers to cases where varied answers are present and the agents fail to reach convergence through the debate. Hence, applying \textit{Early Termination} may result in stopping the debate before the disagreement is resolved, thereby passing the unresolved disagreement to the judge agent, leading to increased difficulty in determining an acceptable answer. Nevertheless, \textit{Extended Reflection} outperforms the SOTA baseline and also improves upon the \textit{Default MAD} on both tasks with statistical significance. This highlights the effectiveness \textit{Extended Reflection} in mitigating \textit{Prolonged Disagreement}, where agent debates involve varied answers and external feedback (i.e., judge agent) enhances the quality of the debate to resolve persistent disagreements.

%NEW IA
This suggests that \textit{Early Termination} was not effective in addressing underperforming cases associated with the \textit{Prolonged Disagreement} pattern (Table~\ref{tab:rq2_categorization}), which appeared most frequently in the Code Input Prediction and Output Prediction tasks. \textit{Prolonged Disagreement} arises when multiple divergent answers are proposed and the debating agents fail to converge. In such cases, applying \textit{Early Termination} risks halting the debate prematurely, thereby passing unresolved disagreements to the judge agent and increasing the difficulty of identifying an acceptable solution. By contrast, \textit{Extended Reflection} not only outperforms the SOTA baseline but also achieves statistically significant improvements over \textit{Default MAD} on both tasks. This underscores the effectiveness of \textit{Extended Reflection} in mitigating \textit{Prolonged Disagreement}, as the incorporation of additional debate rounds and external feedback from the judge agent enables agents to resolve persistent disagreements and reach higher-quality outcomes.

For Code Translation, the \textit{Early Termination} and \textit{Extended Reflection} techniques outperformed the SOTA Baseline, PLTranslation~\cite{pan2024lost} on all three datasets across the two programming languages. The results show that the two enhancement strategies generally outperform the SOTA baseline with statistical significance and medium to large effect sizes. For example, the \textit{Extended Reflection} on Evalplus dataset outperformed SOTA baseline with a large effect size (Qwen for Java Cohen's $d$ = 1.75), and the \textit{Early Termination} on the AVATAR dataset outperformed the SOTA baseline with a medium effect size (Qwen for Python Cohen's $d$ = 0.44). 

% Table~\ref{tab:rq3_winner} shows that for the Code Translation task, \textit{Extended Reflection} achieved the best performance overall.

% two MAD variants vs Default MAD 
%NEW IA
The results also show that both enhancement strategies significantly improved the performance of Default MAD, which by itself did not surpass the SOTA baseline. This indicates that \textit{Early Termination} and \textit{Extended Reflection} are effective in mitigating underperforming cases associated with the \textit{Ending Divergence} pattern (Table~\ref{tab:rq2_categorization}), the most prevalent issue in Code Translation. Specifically, \textit{Early Termination} prevents unnecessary divergence by concluding the debate once an acceptable response emerges, thereby avoiding the propagation of incorrect outputs. In contrast, when divergence occurs, \textit{Extended Reflection} enables agents to restart the debate with feedback from the judge agent, steering the discussion back toward an optimal trajectory.

%The results also show that the two enhancement strategies significantly improved the performance of Default MAD, which did not outperform the SOTA baseline. This suggests that both \textit{Early Termination} and \textit{Extended Reflection} are effective in mitigating underperforming cases associated with the \textit{Ending Divergence} pattern (Table~\ref{tab:rq2_categorization}), which was the most prevalent in Code Translation. Specifically, \textit{Early Termination} avoids unnecessary divergence by ending the debate early once an acceptable responses emerges, avoiding propagation of incorrect responses. In case of divergence, the \textit{Extended Reflection} allows agents to restart the debate with a feedback from the judge agent, guiding the debate back to an optimal trajectory. 

% Overall, the two enhancement strategies significantly improve the performance of MAD on tasks where underperformance is caused by a divergence from correct consensus, by terminating early or steering the debate back toward the correct trajectory.

% (i.e., input and output values for CIP and COP, summary for CS, and translated code for CT).

% CS: two MAD variants vs SOTA

%For all the metrics assessing the textual and semantic similarity between generated and human-written summaries (Section~\ref{cs_metrics}), \textit{Early Termination} and \textit{Extended Reflection} generally outperform the SOTA baseline with statistical significance. 
For the Code Summarization, our two enhancement techniques show superior performance compared to the SOTA baseline, ASCS~\cite{sun2024source}. The \textit{Early Termination} technique outperforms the SOTA baseline with a medium effect size (GPT-4o for Python using BERTScore, Cohen's $d$ = 0.55), and the \textit{Extended Reflection} technique outperforms the SOTA baseline with a medium effect size (CodeLLaMA for Python using USECS, Cohen's $d$ = 0.69). In addition, in terms of the SIDE score, both techniques show superior performance compared to the SOTA baseline.

% two MAD variants vs Default MAD
Further, \textit{Early Termination} and \textit{Extended Reflection} techniques also show improvement compared to the Default MAD. %Similar to Code Translation, this indicates the effectiveness of our two enhancement strategies in improving the performance of the Code Summarization task. 
Since \textit{Early Termination} would stop the debate from digressing from the acceptable response that may be generated early on in the debate process, and the feedback provided by the judge agent in \textit{Extended Reflection} would also help the debate to converge to an acceptable answer, they might contribute to mitigating the \textit{Ending Divergence} that is the most prevalent pattern responsible for underperforming cases for Code Summarization (Table \ref{tab:rq2_categorization}). 

% Overall, our two MAD enhancements exhibit superior performance compared to the SOTA baseline and also improve upon the performance of Default MAD, highlighting the effectiveness of MAD on the Code Summarization task.

\captionsetup{skip=1pt}
\setlength{\tabcolsep}{3pt}
\begin{table}[h]
% \footnotesize
\centering
\caption{
\centering
Summary of Best-Performing Methods 
\\ Based on Results in Table 3, 4, and 5 (per dataset)}
\label{tab:rq3_winner}
\begin{tabular}{lcccc}
\toprule
\textbf{Method} & \textbf{CIP} & \textbf{COP} & \textbf{CT} & \textbf{CS} \\
\midrule
CoT (SA Baseline) & 0 & 0 & 0 & 0 \\
SC (SA Baseline)  & 0 & 0 & 0 & 0 \\
SOTA (SE Baseline) & 0 & 1 & 0 & 0 \\
\midrule
Default MAD                         & 0 & 0 & 0 & 0 \\
Early Termination                   & 0 & 1 & 2 & 1 \\
Extended Reflection                 &\textbf{2} & \textbf{2} &\textbf{10} &\textbf{2} \\
\bottomrule
\end{tabular}
\end{table}

% ── Table A: Single-Agent ──────────────────────────────────────────────────────
\setlength{\tabcolsep}{3pt}
\begin{table*}[hbtp]
\centering
\scriptsize
\caption{Single-Agent Token Usage, API Call Counts, and Runtime}
\resizebox{0.95\linewidth}{!}{
\renewcommand{\arraystretch}{0.5}
\begin{tabular}{llccccccccc}
    \toprule
    \textbf{Task} & \multicolumn{1}{c}{\textbf{Model}} &
    \multicolumn{3}{c}{\textbf{Chain-of-Thought}} &
    \multicolumn{3}{c}{\textbf{Self-Consistency}} &
    \multicolumn{3}{c}{\textbf{SE Baseline}} \\
    \cmidrule(lr){3-5} \cmidrule(lr){6-8} \cmidrule(lr){9-11}
    & & \textbf{API} & \textbf{Tokens} & \textbf{Avg Time (s)}
      & \textbf{API} & \textbf{Tokens} & \textbf{Avg Time (s)}
      & \textbf{API} & \textbf{Tokens} & \textbf{Avg Time (s)} \\
    \midrule
    CIP & Deepseek Coder & \multirow{2}{*}{800} & 436,801 & 10.67 & \multirow{2}{*}{4,000} & 1,444,971 & 37.10 & \multirow{2}{*}{800} & 949,691 & 12.73 \\
    \addlinespace[1pt]
        & GPT-4o         &                    & 493,620 & 5.99 &                    & 1,757,472 &  25.48 &                    & 1,043,398 & 11.46 \\
    \midrule
    COP & Deepseek Coder & \multirow{2}{*}{1,279} & 921,666 & 16.14 & \multirow{2}{*}{6,395} & 3,000,051 & 44.80 & \multirow{2}{*}{1,498} & 1,873,199 & 13.10 \\
    \addlinespace[1pt]
        & GPT-4o         &                    & 840,332 & 6.70 &                    & 3,272,498 & 21.52 &                    & 1,991,383 & 12.90 \\
    \midrule
    CT  & Qwen2.5 Coder  & \multirow{2}{*}{1,227} & 1,108,683 & 19.50 & \multirow{2}{*}{6,135} & 4,342,271 & 70.38 & 5,944 & 3,678.192 & 57.37 \\
    \addlinespace[1pt]
        & GPT-4o         &                    & 1,049,093 & 8.75 &                    & 4,720,204 & 23.60 & 5,304 & 3,410,449 & 29.29 \\
    \midrule
    CS  & CodeLlama 7B   & \multirow{2}{*}{1,330} & 486,954 & 4.91 & \multirow{2}{*}{6,650} & 2,175,641 & 17.68 & \multirow{2}{*}{1,330} & 687,740 & 9.79 \\
    \addlinespace[1pt]
        & GPT-4o         &                    & 541,693 & 3.53 &                    & 2,193,625 &  8.55 &                    & 519,541 & 5.26 \\
    \bottomrule
\end{tabular}
}
\label{tab:sa_token_usage}
\end{table*}

% ── Table B: Multi-Agent ───────────────────────────────────────────────────────
\setlength{\tabcolsep}{2pt}
\begin{table*}[htbp]
\centering
\caption{\centering Multi-Agent Token Usage, API Call Counts, and Runtime \\
\scriptsize\textit{($+$) and ($-$) indicate higher or lower API call counts and token usage, relative to Default MAD}}
\resizebox{0.95\linewidth}{!}{
\begin{tabular}{llccccccccc}
    \toprule
    \textbf{Task} & \multicolumn{1}{c}{\textbf{Model}} &
    \multicolumn{3}{c}{\textbf{Default MAD}} &
    \multicolumn{3}{c}{\textbf{Early Termination}} &
    \multicolumn{3}{c}{\textbf{Extended Reflection}} \\
    \cmidrule(lr){3-5} \cmidrule(lr){6-8} \cmidrule(lr){9-11}
    & & \textbf{API} & \textbf{Tokens} & \textbf{Avg Time (s)}
      & \textbf{API} & \textbf{Tokens} & \textbf{Avg Time (s)}
      & \textbf{API} & \textbf{Tokens} & \textbf{Avg Time (s)} \\
    \midrule
    CIP & Deepseek Coder & \multirow{2}{*}{8,000} & 22,031,559 & 16.37 & 11,544 \textbf{(+)} & 18,132,408 \textbf{(--)} & 39.10 \textbf{(+)} & 10,333 \textbf{(+)} & 19,595,300 \textbf{(--)} & 35.61 \textbf{(+)} \\
    \addlinespace[1pt]
        & GPT-4o         &                         & 37,686,633 & 9.23 & 14,545 \textbf{(+)} & 24,053,281 \textbf{(--)} & 25.41 \textbf{(+)} & 9,657 \textbf{(+)}  & 26,879,949 \textbf{(--)} & 24.48 \textbf{(+)} \\
    \midrule
    COP & Deepseek Coder & \multirow{2}{*}{12,790} & 15,910,258 & 21.83 & 14,862 \textbf{(+)} & 12,318,309 \textbf{(--)} & 37.46 \textbf{(+)} & 18,422 \textbf{(+)} & 12,786,392 \textbf{(--)} & 37.02 \textbf{(+)} \\
    \addlinespace[1pt]
        & GPT-4o         &                          & 29,453,770 & 9.19 & 11,081 \textbf{(--)} & 21,745,744 \textbf{(--)} & 20.60 \textbf{(+)} & 18,511 \textbf{(+)} & 24,740,981 \textbf{(--)} & 22.79 \textbf{(+)} \\
    \midrule
    CT  & Qwen2.5 Coder  & \multirow{2}{*}{12,020} & 23,822,582 & 30.03 & 9,644 \textbf{(--)}  & 19,675,294 \textbf{(--)} & 47.03 \textbf{(+)} & 12,916 \textbf{(+)} & 20,745,873 \textbf{(--)} & 39.86 \textbf{(+)} \\
    \addlinespace[1pt]
        & GPT-4o         &                          & 25,023,335 & 23.30 & 8,381 \textbf{(--)}  & 21,865,206 \textbf{(--)} & 34.10 \textbf{(+)} & 11,376 \textbf{(+)} & 22,033,307 \textbf{(--)} & 30.54 \textbf{(+)} \\
    \midrule
    CS  & CodeLlama 7B   & \multirow{2}{*}{13,300} & 22,791,951 & 34.53 & 15,950 \textbf{(+)} & 16,911,518 \textbf{(--)} & 29.68 \textbf{(--)} & 15,952 \textbf{(+)} & 21,072,013 \textbf{(--)} & 24.32 \textbf{(--)} \\
    \addlinespace[1pt]
        & GPT-4o         &                          & 23,379,922 & 23.78 & 15,155 \textbf{(+)} & 18,983,360 \textbf{(--)} & 20.49 \textbf{(--)} & 14,417 \textbf{(+)} & 22,955,734 \textbf{(--)} & 20.66 \textbf{(--)} \\
    \bottomrule
\end{tabular}
}
\label{tab:mad_token_usage}
\end{table*}

In addition, Table~\ref{tab:rq3_winner} shows the count of best-performing techniques per task based on model–dataset combinations. For example, the value 2 in the CIP column for \textit{Extended Reflection} indicates that this technique achieved the highest performance on the CRUXEval dataset using two models (DeepSeek and GPT-4o) for the Code Input Prediction task. Our results suggest that \textit{Extended Reflection} consistently achieves the best performance across the coding tasks, datasets, and models used in our experiments (Section~\ref{sec:baselines}).

% \input{tables/model_token_usage}

% para 9

Tables~\ref{tab:sa_token_usage} and~\ref{tab:mad_token_usage} report the total number of API calls, total token consumption, and average execution time per sample for the Single-Agent baselines and Multi-Agent methods. In Table~\ref{tab:mad_token_usage}, \textbf{(--)} denotes a decrease and \textbf{(+)} denotes an increase relative to Default MAD. As shown in Tables~\ref{tab:sa_token_usage} and~\ref{tab:mad_token_usage}, single-agent approaches generally exhibit lower token usage, fewer API calls, and shorter execution times than MAD. As multi-agent systems inherently involve multiple LLM invocations, improving reasoning performance introduces additional computational overhead. These observations reinforce the need for more structured and efficient coordination mechanisms, such as MAD, to better utilize computational resources.

Notably, both MAD variants consistently reduce token consumption across all four coding tasks compared to Default MAD, while API call counts and average execution times vary across tasks and models. This variation can be attributed to the increased involvement of the judge agent, which evaluates early termination conditions in the \textit{Early Termination} variant and triggers debate restarts in the \textit{Extended Reflection} variant. Overall, these results suggest that Default MAD tends to incur unnecessary token overhead by prolonging debates beyond optimal trajectories or inducing verbosity in agent responses even after a valid answer has emerged, thereby wasting computational resources and potentially degrading output quality.

\begin{tcolorbox}[
  colback=white,
  boxsep=1pt,       % tighter internal padding
  left=2pt,         % left padding
  right=2pt,        % right padding
  top=2pt,          % top padding
  bottom=2pt,       % bottom padding
]
\textbf{Observation 3:} Our two enhancement improve performance of MAD on all four coding tasks with \textit{Extended Reflection} strategy the best-performance overall.
\end{tcolorbox}

To further quantify this trade off, Table \ref{tab:roi_cip}, \ref{tab:roi_cop}, \ref{tab:roi_ct}, and \ref{tab:roi_cs} report the Return-on-Investment (ROI) computed as the difference in performance result over each measure of resource consumption, including API calls, token consumption, and average execution time. For example, the values in the first row of Table~\ref{tab:roi_cip} are computed by comparing Default MAD with the CoT baseline using DeepSeek-Coder. The absolute performance difference (Abs$\Delta$) is computed as the difference in pass@1 scores between Default MAD and CoT (31.9 - 31.8 = 0.10), based on Table~\ref{tab:cr_io_mad}. The relative performance difference (Rel$\Delta$\%) is calculated by normalizing this difference by the baseline score (0.10 / 31.8 $\times$ 100 = 0.31\%). The ROI per second (ROI/s) is then computed by dividing the absolute performance difference by the difference in average time per sample (16.37 - 10.67 = 5.70s), yielding an ROI/s of 0.02. For Code Summarization, which is evaluated across multiple metrics, the performance difference is first averaged across all metrics prior to ROI computation. Additionally, ROI for token consumption is reported per million tokens to improve readability.

A positive ROI indicates that the additional computational cost incurred by MAD (in terms of tokens, API calls, or execution time) is justified by the performance gain over the baseline. Among the MAD variants, \textit{Extended Reflection} demonstrates consistently high ROI when compared against Single-Agent baselines, achieving a positive ROI across all tasks. Notably, for Code Summarization, all MAD frameworks yield a positive ROI compared to the Single-Agent baselines, suggesting a clear cost-benefit advantage of adopting MAD for this task. For Code Translation, our proposed MAD variants similarly demonstrate a positive ROI overall, with a particular advantage in overcoming the complete failure (pass@1 = 0\% for Python to Java translation cases) that single-agent approaches are unable to resolve. In contrast, the computational overhead may not always translate into performance improvements as the \textit{Default MAD} showed lower or comparable performance to the SOTA SE technique, resulting in a negative ROI for Code Input and Output Prediction tasks.

% Requires: \usepackage{booktabs,multirow,xcolor}
% Paste each table block into your .tex file, or \input this file.

% ---- Task: CIP ----
\begin{table*}[htbp]
  \centering
  \captionsetup{skip=1pt}
  \caption{\centering ROI Analysis for CIP\\
  \scriptsize $\uparrow$: MAD outperforms single-agent baseline; $\downarrow$: underperforms}
  \label{tab:roi_cip}
  \resizebox{0.95\linewidth}{!}{
  \begin{tabular}{ll r r r r r r r r r r r r r r}
    \toprule
    & & \multicolumn{7}{c}{DeepSeek} &\multicolumn{7}{c}{GPT-4o} \\
    \cmidrule(lr){3-9}\cmidrule(lr){10-16}
    \textbf{MAD Variant} & \textbf{Methods} & \scriptsize Abs$\Delta$ & \scriptsize Rel$\Delta$(\%) & \scriptsize ROI/MTok & \scriptsize ROI/API & \scriptsize ROI/s & \scriptsize Avg ROI & \scriptsize $\uparrow$ / $\downarrow$ Perf & \scriptsize Abs$\Delta$ & \scriptsize Rel$\Delta$(\%) & \scriptsize ROI/MTok & \scriptsize ROI/API & \scriptsize ROI/s & \scriptsize Avg ROI & \scriptsize $\uparrow$ / $\downarrow$ Perf \\
    \midrule
    \multirow{3}{*}{\textit{Default}} & CoT & 0.10 & 0.31 & 0.00 & 0.00 & 0.02 & 0.01 & $\uparrow$ & 21.90 & 63.48 & 0.59 & 0.30 & 6.76 & 2.55 & $\uparrow$ \\
     & Self-Con. & -1.70 & -5.06 & -0.08 & -0.04 & -0.08 & -0.07 & $\downarrow$ & 21.70 & 62.54 & 0.60 & 0.54 & 1.34 & 0.83 & $\uparrow$ \\
     & SE & -30.60 & -48.96 & -1.45 & -0.43 & -8.41 & -3.43 & $\downarrow$ & -12.90 & -18.61 & -0.35 & -0.18 & -5.78 & -2.11 & $\downarrow$ \\
    \midrule
    \multirow{3}{*}{\textit{Early Term.}} & CoT & 10.80 & 33.96 & 0.61 & 0.10 & 0.38 & 0.36 & $\uparrow$ & 12.30 & 35.65 & 0.52 & 0.09 & 0.63 & 0.41 & $\uparrow$ \\
     & Self-Con. & 9.00 & 26.79 & 0.54 & 0.12 & 4.50 & 1.72 & $\uparrow$ & 12.10 & 34.87 & 0.54 & 0.11 & 172.86 & 57.84 & $\uparrow$ \\
     & SE & -19.90 & -31.84 & -1.16 & -0.19 & -0.75 & -0.70 & $\downarrow$ & -22.50 & -32.47 & -0.98 & -0.16 & -1.61 & -0.92 & $\downarrow$ \\
    \midrule
    \multirow{3}{*}{\textit{Ext. Reflect.}} & CoT & 32.70 & 102.83 & 1.71 & 0.34 & 1.31 & 1.12 & $\uparrow$ & 39.00 & 113.04 & 1.48 & 0.44 & 2.11 & 1.34 & $\uparrow$ \\
     & Self-Con. & 30.90 & 91.96 & 1.70 & 0.49 & 20.74 & 7.64 & $\uparrow$ & 38.80 & 111.82 & 1.54 & 0.69 & 38.80 & 13.68 & $\uparrow$ \\
     & SE & 2.00 & 3.20 & 0.11 & 0.02 & 0.09 & 0.07 & $\uparrow$ & 4.20 & 6.06 & 0.16 & 0.05 & 0.32 & 0.18 & $\uparrow$ \\
    \bottomrule
  \end{tabular}
  }
\end{table*}

% ---- Task: COP ----
\begin{table*}[htbp]
  \centering
  \captionsetup{skip=1pt}
  \caption{\centering ROI Analysis for COP\\
  \scriptsize Averaged across CRUXEval and LiveCodeBench. $\uparrow$: MAD outperforms single-agent baseline; $\downarrow$: underperforms}
  \label{tab:roi_cop}
  \resizebox{0.95\linewidth}{!}{
  \begin{tabular}{ll r r r r r r r r r r r r r r}
    \toprule
    & & \multicolumn{7}{c}{DeepSeek} &\multicolumn{7}{c}{GPT-4o} \\
    \cmidrule(lr){3-9}\cmidrule(lr){10-16}
    \textbf{MAD Variant} & \textbf{Methods} & \scriptsize Abs$\Delta$ & \scriptsize Rel$\Delta$(\%) & \scriptsize ROI/MTok & \scriptsize ROI/API & \scriptsize ROI/s & \scriptsize Avg ROI & \scriptsize $\uparrow$ / $\downarrow$ Perf & \scriptsize Abs$\Delta$ & \scriptsize Rel$\Delta$(\%) & \scriptsize ROI/MTok & \scriptsize ROI/API & \scriptsize ROI/s & \scriptsize Avg ROI & \scriptsize $\uparrow$ / $\downarrow$ Perf \\
    \midrule
    \multirow{3}{*}{\textit{Default}} & CoT & 15.00 & 54.25 & 1.00 & 0.13 & 2.64 & 1.26 & $\uparrow$ & 26.10 & 203.11 & 0.91 & 0.23 & 10.48 & 3.87 & $\uparrow$ \\
     & Self-Con. & -8.25 & -16.21 & -0.64 & -0.13 & -0.36 & -0.38 & $\downarrow$ & 26.65 & 216.67 & 1.02 & 0.42 & 2.16 & 1.20 & $\uparrow$ \\
     & SE & -19.75 & -31.65 & -1.41 & -0.17 & -2.26 & -1.28 & $\downarrow$ & -30.50 & -43.92 & -1.11 & -0.27 & -8.22 & -3.20 & $\downarrow$ \\
    \midrule
    \multirow{3}{*}{\textit{Early Term.}} & CoT & 45.95 & 166.18 & 4.03 & 0.34 & 2.16 & 2.18 & $\uparrow$ & 39.45 & 307.00 & 1.89 & 0.40 & 2.84 & 1.71 & $\uparrow$ \\
     & Self-Con. & 22.70 & 44.60 & 2.44 & 0.27 & 3.09 & 1.93 & $\uparrow$ & 40.00 & 325.20 & 2.17 & 0.85 & 43.48 & 15.50 & $\uparrow$ \\
     & SE & 11.20 & 17.95 & 1.07 & 0.08 & 0.46 & 0.54 & $\uparrow$ & -17.15 & -24.69 & -0.87 & -0.18 & -2.23 & -1.09 & $\downarrow$ \\
    \midrule
    \multirow{3}{*}{\textit{Ext. Reflect.}} & CoT & 48.20 & 174.32 & 4.06 & 0.28 & 2.31 & 2.22 & $\uparrow$ & 57.95 & 450.97 & 2.42 & 0.34 & 3.60 & 2.12 & $\uparrow$ \\
     & Self-Con. & 24.95 & 49.02 & 2.55 & 0.21 & 3.21 & 1.99 & $\uparrow$ & 58.50 & 475.61 & 2.72 & 0.48 & 46.06 & 16.42 & $\uparrow$ \\
     & SE & 13.45 & 21.55 & 1.23 & 0.08 & 0.56 & 0.62 & $\uparrow$ & 1.35 & 1.94 & 0.06 & 0.01 & 0.14 & 0.07 & $\uparrow$ \\
    \bottomrule
  \end{tabular}
  }
\end{table*}

% ---- Task: CT ----
\begin{table*}[htbp]
  \centering
  \captionsetup{skip=1pt}
  \caption{\centering ROI Analysis for CT\\
  \scriptsize $\uparrow$: MAD outperforms single-agent baseline; $\downarrow$: underperforms}
  \label{tab:roi_ct}
  \resizebox{0.95\linewidth}{!}{
  \begin{tabular}{lll r r r r r r r r r r r r r r}
    \toprule
    & & & \multicolumn{7}{c}{Qwen2.5} &\multicolumn{7}{c}{GPT-4o} \\
    \cmidrule(lr){4-10}\cmidrule(lr){11-17}
    \textbf{MAD Variant} & \textbf{Methods} & \textbf{Lang} & \scriptsize Abs$\Delta$ & \scriptsize Rel$\Delta$(\%) & \scriptsize ROI/MTok & \scriptsize ROI/API & \scriptsize ROI/s & \scriptsize Avg ROI & \scriptsize $\uparrow$ / $\downarrow$ Perf & \scriptsize Abs$\Delta$ & \scriptsize Rel$\Delta$(\%) & \scriptsize ROI/MTok & \scriptsize ROI/API & \scriptsize ROI/s & \scriptsize Avg ROI & \scriptsize $\uparrow$ / $\downarrow$ Perf \\
    \midrule
    \multirow{6}{*}{\textit{Default}} & \multirow{2}{*}{CoT} & Python & -5.07 & -9.78 & -0.22 & -0.05 & -0.48 & -0.25 & $\downarrow$ & -17.86 & -33.41 & -0.74 & -0.17 & -1.23 & -0.71 & $\downarrow$ \\
     &  & Java & 51.94 & -- & 2.29 & 0.48 & 4.93 & 2.57 & $\uparrow$ & 38.55 & -- & 1.61 & 0.36 & 2.65 & 1.54 & $\uparrow$ \\
     & \multirow{2}{*}{Self-Con.} & Python & -10.90 & -18.90 & -0.56 & -0.19 & -0.27 & -0.34 & $\downarrow$ & -15.20 & -29.92 & -0.75 & -0.26 & -50.67 & -17.22 & $\downarrow$ \\
     &  & Java & 51.94 & -- & 2.67 & 0.88 & 1.29 & 1.61 & $\uparrow$ & 38.55 & -- & 1.90 & 0.66 & 128.50 & 43.68 & $\uparrow$ \\
     & \multirow{2}{*}{SE} & Python & 0.55 & 1.19 & 0.03 & 0.01 & 0.02 & 0.02 & $\uparrow$ & -1.93 & -5.14 & -0.09 & -0.03 & -0.32 & -0.15 & $\downarrow$ \\
     &  & Java & 2.18 & 4.38 & 0.11 & 0.04 & 0.08 & 0.07 & $\uparrow$ & 11.62 & 43.15 & 0.54 & 0.17 & 1.94 & 0.88 & $\uparrow$ \\
    \midrule
    \multirow{6}{*}{\textit{Early Term.}} & \multirow{2}{*}{CoT} & Python & 11.09 & 21.39 & 0.60 & 0.13 & 0.40 & 0.38 & $\uparrow$ & -12.73 & -23.81 & -0.61 & -0.18 & -0.50 & -0.43 & $\downarrow$ \\
     &  & Java & 65.58 & -- & 3.53 & 0.78 & 2.38 & 2.23 & $\uparrow$ & 47.67 & -- & 2.29 & 0.67 & 1.88 & 1.61 & $\uparrow$ \\
     & \multirow{2}{*}{Self-Con.} & Python & 5.26 & 9.12 & 0.34 & 0.15 & 0.23 & 0.24 & $\uparrow$ & -10.07 & -19.82 & -0.59 & -0.45 & -0.96 & -0.66 & $\downarrow$ \\
     &  & Java & 65.58 & -- & 4.28 & 1.87 & 2.81 & 2.98 & $\uparrow$ & 47.67 & -- & 2.78 & 2.12 & 4.54 & 3.15 & $\uparrow$ \\
     & \multirow{2}{*}{SE} & Python & 16.71 & 36.15 & 1.04 & 0.45 & 1.62 & 1.04 & $\uparrow$ & 3.20 & 8.53 & 0.17 & 0.10 & 0.67 & 0.31 & $\uparrow$ \\
     &  & Java & 15.82 & 31.79 & 0.99 & 0.43 & 1.53 & 0.98 & $\uparrow$ & 20.74 & 77.01 & 1.12 & 0.67 & 4.31 & 2.04 & $\uparrow$ \\
    \midrule
    \multirow{6}{*}{\textit{Ext. Reflect.}} & \multirow{2}{*}{CoT} & Python & 18.14 & 34.99 & 0.92 & 0.16 & 0.89 & 0.66 & $\uparrow$ & -3.51 & -6.57 & -0.17 & -0.03 & -0.16 & -0.12 & $\downarrow$ \\
     &  & Java & 75.94 & -- & 3.87 & 0.65 & 3.73 & 2.75 & $\uparrow$ & 65.45 & -- & 3.12 & 0.64 & 3.00 & 2.26 & $\uparrow$ \\
     & \multirow{2}{*}{Self-Con.} & Python & 12.31 & 21.35 & 0.75 & 0.18 & 0.40 & 0.45 & $\uparrow$ & -0.85 & -1.67 & -0.05 & -0.02 & -0.12 & -0.06 & $\downarrow$ \\
     &  & Java & 75.94 & -- & 4.63 & 1.12 & 2.49 & 2.75 & $\uparrow$ & 65.45 & -- & 3.78 & 1.25 & 9.43 & 4.82 & $\uparrow$ \\
     & \multirow{2}{*}{SE} & Python & 23.76 & 51.41 & 1.39 & 0.34 & 1.36 & 1.03 & $\uparrow$ & 12.42 & 33.09 & 0.67 & 0.20 & 9.94 & 3.60 & $\uparrow$ \\
     &  & Java & 26.18 & 52.61 & 1.53 & 0.38 & 1.50 & 1.13 & $\uparrow$ & 38.52 & 143.04 & 2.07 & 0.63 & 30.82 & 11.17 & $\uparrow$ \\
    \bottomrule
  \end{tabular}
  }
\end{table*}

% ---- Task: CS ----
\begin{table*}[htbp]
  \centering
  \captionsetup{skip=1pt}
  \caption{\centering ROI Analysis for CS\\
  \scriptsize $\uparrow$: MAD outperforms single-agent baseline; $\downarrow$: underperforms}
  \label{tab:roi_cs}
  \resizebox{0.95\linewidth}{!}{
  \begin{tabular}{lll r r r r r r r r r r r r r r}
    \toprule
    & & & \multicolumn{7}{c}{CodeLLaMA} &\multicolumn{7}{c}{GPT-4o} \\
    \cmidrule(lr){4-10}\cmidrule(lr){11-17}
    \textbf{MAD Variant} & \textbf{Methods} & \textbf{Lang} & \scriptsize Abs$\Delta$ & \scriptsize Rel$\Delta$(\%) & \scriptsize ROI/MTok & \scriptsize ROI/API & \scriptsize ROI/s & \scriptsize Avg ROI & \scriptsize $\uparrow$ / $\downarrow$ Perf & \scriptsize Abs$\Delta$ & \scriptsize Rel$\Delta$(\%) & \scriptsize ROI/MTok & \scriptsize ROI/API & \scriptsize ROI/s & \scriptsize Avg ROI & \scriptsize $\uparrow$ / $\downarrow$ Perf \\
    \midrule
    \multirow{6}{*}{\textit{Default}} & \multirow{2}{*}{CoT} & Python & 6.62 & 97.78 & 0.30 & 0.06 & 0.22 & 0.19 & $\uparrow$ & 11.46 & 163.25 & 0.50 & 0.10 & 0.57 & 0.39 & $\uparrow$ \\
     &  & Java & 7.50 & 109.97 & 0.34 & 0.06 & 0.25 & 0.22 & $\uparrow$ & 9.43 & 138.47 & 0.41 & 0.08 & 0.47 & 0.32 & $\uparrow$ \\
     & \multirow{2}{*}{Self-Con.} & Python & 6.68 & 99.55 & 0.32 & 0.10 & 0.40 & 0.27 & $\uparrow$ & 12.19 & 193.80 & 0.58 & 0.18 & 0.80 & 0.52 & $\uparrow$ \\
     &  & Java & 7.25 & 102.55 & 0.35 & 0.11 & 0.43 & 0.30 & $\uparrow$ & 9.51 & 141.31 & 0.45 & 0.14 & 0.62 & 0.41 & $\uparrow$ \\
     & \multirow{2}{*}{SE} & Python & 4.20 & 45.70 & 0.19 & 0.04 & 0.17 & 0.13 & $\uparrow$ & 2.68 & 16.96 & 0.12 & 0.02 & 0.14 & 0.09 & $\uparrow$ \\
     &  & Java & 1.32 & 10.15 & 0.06 & 0.01 & 0.05 & 0.04 & $\uparrow$ & 1.41 & 9.51 & 0.06 & 0.01 & 0.08 & 0.05 & $\uparrow$ \\
    \midrule
    \multirow{6}{*}{\textit{Early Term.}} & \multirow{2}{*}{CoT} & Python & 6.97 & 102.95 & 0.42 & 0.05 & 0.28 & 0.25 & $\uparrow$ & 14.65 & 208.69 & 0.79 & 0.11 & 0.86 & 0.59 & $\uparrow$ \\
     &  & Java & 7.52 & 110.26 & 0.46 & 0.05 & 0.30 & 0.27 & $\uparrow$ & 13.52 & 198.53 & 0.73 & 0.10 & 0.80 & 0.54 & $\uparrow$ \\
     & \multirow{2}{*}{Self-Con.} & Python & 7.03 & 104.77 & 0.48 & 0.08 & 0.59 & 0.38 & $\uparrow$ & 15.38 & 244.52 & 0.92 & 0.18 & 1.29 & 0.79 & $\uparrow$ \\
     &  & Java & 7.27 & 102.83 & 0.49 & 0.08 & 0.61 & 0.39 & $\uparrow$ & 13.60 & 202.08 & 0.81 & 0.16 & 1.14 & 0.70 & $\uparrow$ \\
     & \multirow{2}{*}{SE} & Python & 4.55 & 49.51 & 0.28 & 0.03 & 0.23 & 0.18 & $\uparrow$ & 5.87 & 37.15 & 0.32 & 0.04 & 0.39 & 0.25 & $\uparrow$ \\
     &  & Java & 1.34 & 10.31 & 0.08 & 0.01 & 0.07 & 0.05 & $\uparrow$ & 5.50 & 37.09 & 0.30 & 0.04 & 0.36 & 0.23 & $\uparrow$ \\
    \midrule
    \multirow{6}{*}{\textit{Ext. Reflect.}} & \multirow{2}{*}{CoT} & Python & 9.55 & 141.06 & 0.46 & 0.07 & 0.49 & 0.34 & $\uparrow$ & 16.85 & 240.03 & 0.75 & 0.13 & 0.98 & 0.62 & $\uparrow$ \\
     &  & Java & 7.05 & 103.37 & 0.34 & 0.05 & 0.36 & 0.25 & $\uparrow$ & 16.36 & 240.23 & 0.73 & 0.12 & 0.96 & 0.60 & $\uparrow$ \\
     & \multirow{2}{*}{Self-Con.} & Python & 9.61 & 143.22 & 0.51 & 0.10 & 1.45 & 0.69 & $\uparrow$ & 17.58 & 279.49 & 0.85 & 0.23 & 1.45 & 0.84 & $\uparrow$ \\
     &  & Java & 6.80 & 96.18 & 0.36 & 0.07 & 1.02 & 0.49 & $\uparrow$ & 16.44 & 244.28 & 0.79 & 0.21 & 1.36 & 0.79 & $\uparrow$ \\
     & \multirow{2}{*}{SE} & Python & 7.13 & 77.58 & 0.35 & 0.05 & 0.49 & 0.30 & $\uparrow$ & 8.07 & 51.08 & 0.36 & 0.06 & 0.52 & 0.32 & $\uparrow$ \\
     &  & Java & 0.87 & 6.69 & 0.04 & 0.01 & 0.06 & 0.04 & $\uparrow$ & 8.34 & 56.24 & 0.37 & 0.06 & 0.54 & 0.33 & $\uparrow$ \\
    \bottomrule
  \end{tabular}
  }
\end{table*}

\section{Discussion}
\label{section:discussion}
% Summarize and highlight the main contribution

% highlight the overall contribution :: para 1
The analysis of debate logs and the underperforming cases suggests that while MAD, designed for general NLP tasks, fosters agent collaboration and enhances synergy among multiple LLMs through iterative refinements and information exchange, certain debate interaction patterns can lead to low-quality responses. For instance, agents may converge on an erroneous answer despite an acceptable response being present early in the debate, premature agreement between agents may result in gradual divergence, or agents may adhere to deep-rooted arguments without considering others' perspectives, restricting the advantages of iterative refinement through debate. Moreover, our manual analysis of each underperforming debate category reveals that the most prevalent root causes are agents failing to challenge shared assumptions, progressively identifying different faults across iterations, and persistently reasserting initial answers without providing additional rationale.
By addressing these limitations, the MAD variants, enhanced by the optimization strategies, significantly outperform SOTA Single-Agent prompting techniques and SOTA SE techniques in four selected coding tasks by generating accurate input and output values, generating better code summaries, and translating code that is functionally aligned with the reference code.

% limitation of MAD and benefical cases of using MAD
% MAD is not a silver bullet :: para 2-3
While the structured debate approach of MAD demonstrates effectiveness for the four coding tasks in our study, it may not be readily applicable to all SE tasks. Our results suggest that multi-agent debate offers a structural advantage that single-agent approaches lack: when an agent commits to an incorrect answer, there is no internal mechanism for recovery or refinement of answers, whereas the diversity of reasoning paths in MAD systems enables agents to collectively escape such local minima. This is most evident in our Code Translation task (Python to Java) in Table~\ref{tab:ct_mad_all}, where single-agent methods frequently failed to produce executable Java code, while MAD methods were able to recover from such failures through iterative argumentation.

However, our findings show that the effectiveness of MAD varies across tasks depending on task characteristics and expected outputs. In underperforming cases, interaction patterns and root causes of failure varied across tasks and debate categories. In addition, our results suggest that consensus among stochastic agents does not guarantee correctness, revealing a fundamental limitation of debate-based approaches. As shown in Table~\ref{tab:root_cause}, agents can converge on wrong answers due to shared hallucination, yielding suboptimal debate outcomes. This non-determinism may pose a practical barrier, particularly for SE tasks that demand exactness of outcome. Our study takes a pioneering step toward understanding the applicability of MAD across four atomic coding tasks, addressing the code understanding and reasoning aspects of software engineering. Further research is needed to systematically identify the characteristics of SE tasks that benefit from MAD integration and those that pose fundamental challenges, enabling a more principled and broader application of debate-based approaches across the SE domain.

% reviewer 3 comment 1: "when to use / when not to use" MAD :: para 4
To summarize, our results suggest that MAD may be particularly beneficial for tasks that involve diverse reasoning paths, where single-agent methods are prone to local errors, where multiple plausible solutions must be evaluated, or where high-quality domain-specific training data is limited. In contrast, MAD may be less suitable for tasks requiring highly deterministic outputs, tasks requiring executable or oracle-based verification mechanisms, or tasks where incorrect consensus among agents may reinforce hallucinations rather than mitigate them.

% Is MAD applicable to other SE tasks? para 5-6
For instance, tasks such as requirements engineering and architectural design may benefit from more extensive exploration of role specialization and expertise prompting, whereas repository-level SE tasks such as real-world bug fixing or code generation may require deeper consideration of context provisioning within the debate history to foster meaningful agent interaction, or oracle-based evaluation to validate intermediate outputs during the debate process.

MAD presents a promising solution for addressing coding tasks constrained by the scarcity of high-quality, domain-specific datasets, which has an impact on the performance of learning-based single LLM techniques. In vulnerability detection, for instance, the prevalence of imbalanced datasets (non-vulnerable samples outnumber vulnerable ones) can significantly impede the effectiveness of LLM fine-tuning~\cite{guo2023empirical, le2024mitigating}. Additionally, in Code Translation tasks for less common programming languages, the absence of multilingual parallel reference code poses a challenge for supervised-learning approaches for LLM techniques~\cite{tao2024unraveling}. In such scenarios, MAD offers a viable alternative.

% In this study, we examined interaction patterns between agents and identified the root causes of failing debate interactions, but other aspects of the debate, such as how agents attend to preceding responses when formulating their answers, could be explored to gain deeper insights into MAD's applicability to other SE tasks. Investigating agent interaction patterns, the applicability of MAD for tasks that demand exactness of outcome, and empirical validation across diverse SE settings remain important directions for future work.

% unchanged para 7
In this study, we examined interaction patterns between agents and identified the root causes of failing debate interactions, but other aspects of the debate, such as how agents attend to preceding responses when formulating their answers, could be explored to gain deeper insights into MAD's applicability to other SE tasks. Investigating agent interaction patterns, the applicability of MAD for tasks that demand exactness of outcome, and empirical validation across diverse SE settings remain important directions for future work.
%
% reviewer 3 comment 2: futurue work direction  :: para 7
%
Furthermore, an implication of our work is the use of MAD as a reliability mechanism that steers agent interactions toward more accurate outcomes through iterative argumentation. While MAD improves reliability through interactions among multiple LLM agents, existing works on neuro-symbolic approaches often incorporate additional sources of evidence, such as symbolic execution, formal specifications, executable tests, or program analysis constraints, to validate or correct LLM-generated outputs.
In this study, we investigate the applicability and effectiveness of MAD on four coding tasks, and a direct comparison between MAD and neuro-symbolic approaches is beyond the scope of this work. Future work could investigate MAD from a reliability perspective to examine whether MAD can be effectively used in conjunction with neuro-symbolic approaches to improve the reliability and robustness of MAD-based systems.

% MAD and ROI analysis:: para 8
Further, despite the performance improvements of the two MAD variants over \textit{Default MAD}, it is important to consider the return on investment (ROI) when employing MAD for coding tasks. Notably, the MAD variants equipped with our optimization strategies drastically reduced the number of API calls across the four selected coding tasks compared to \textit{Default MAD}, yet the total number of inferences required to generate the final answer remains significantly large compared to single LLM-based approaches. Therefore, the adoption of MAD for end-to-end task automation is most compelling when performance gains outweigh the additional computational cost, including tasks where single-agent approaches consistently fail to produce acceptable outputs, tasks that require iterative validation of intermediate outputs, and tasks where relying on a single agent introduces risk and necessitates a distribution of responsibility among multiple solving agents. Identifying criteria for determining when to employ single-agent versus multi-agent configurations, and leveraging such criteria to maximize performance and computational efficiency, remains an important open question for the research community.

\section{Threats to Validity}
\label{section:threats}
%We took deliberate steps to minimize the potential threats to the validity of this study. In this section, we will discuss these threats.

\noindent\textbf{Construct validity}
The quality of predicted input and output of a program, translated code, and generated summaries for the selected coding tasks (Section~\ref{section:related work}) is influenced by the prompts used for MAD agents. We adopted prompting strategies from prior NLP MAD studies~\cite{liang-etal-2024-encouraging, du2023improving, chan2024chateval}, ensuring that agents engage in effective debate behaviors and maintain high-quality outputs.

To mitigate bias in manual evaluation, two authors independently analyzed debate logs to address potential threats in categorizing debate interaction patterns. Disagreements were resolved through a negotiated agreement. Additionally, since the prompts could influence LLM-generated annotations, we followed established prompting practices for data annotation. We observed strong alignment between LLM annotations and human categorizations, further reducing potential bias. The debate interaction patterns observed in underperforming debates do not encompass all possible forms of misbehavior or misalignment in MAD discussions, and the impact of debate patterns not discovered in our study remains unknown.

\noindent\textbf{Internal Validity} 
The selection of the MAD framework by Liang et al.~\cite{liang-etal-2024-encouraging} as our baseline may influence our findings, as alternative MAD implementations such as those proposed by Du et al.~\cite{du2023improving} or ChatEval~\cite{chan2024chateval} could yield different results. While we selected Liang et al.~\cite{liang-etal-2024-encouraging} as the most recent and task-relevant framework, we acknowledge that conclusions about MAD's effectiveness on coding tasks may not generalize across all MAD design choices.

To ensure that observed improvements are attributable to the MAD architecture rather than variations in LLM inference parameters, we maintained consistent debate configurations and inference settings across all experiments, as detailed in Section~\ref{method_debate_config}. Additionally, as described in Section~\ref{sec:baselines}, we used the same model and dataset to compare MAD against SOTA. While the dataset may be susceptible to data leakage, this limitation applies equally to all compared methods; thus, our analysis focuses on relative performance differences to assess the advantages of MAD.

\noindent\textbf{External Validity}
Our study focuses on six datasets covering four specific coding tasks in Python and Java, which are file-level, atomic tasks addressing the code understanding and reasoning aspects of software engineering. For our selected tasks and datasets, we use three open-source and one proprietary LLM with varying parameter sizes, which may limit the generalizability of our results to other software engineering tasks or programming languages.

\section{Conclusion}
\label{section:conclusion}
This study explored the effectiveness of MAD on four coding tasks, bridging the gap between MAD in NLP and its application to software engineering. Our findings lay the groundwork for future research on harnessing debate-driven reasoning and collaborative synergy in SE contexts. Future work should investigate additional coding tasks suited for MAD, analyze debate interaction patterns to build a taxonomy of agent interactions and assess the cost-effectiveness of MAD compared to other approaches requiring multiple LLM inferences.

\section{DATA AVAILABILITY}
\label{section:data}

As part of our commitment to open science policy, all data collected for this study are made available as supplemental material. We provide our replication package in~\cite{supply}.

\newpage
\bibliographystyle{ACM-Reference-Format}
\bibliography{refs}

\end{document}